\documentclass[twocolumn,aps,showpacs]{revtex4-1}
\usepackage[latin9]{inputenc}
\usepackage{array}
\usepackage{textcomp}
\usepackage{amsmath}
\usepackage{amssymb}
\usepackage{graphicx}
\usepackage{subscript}

\makeatletter

\InputIfFileExists{t2aenc.def}{}{%
  \errmessage{File `t2aenc.def' not found: Cyrillic script not supported}}
\DeclareRobustCommand{\cyrtext}{%
  \fontencoding{T2A}\selectfont\def\encodingdefault{T2A}}
\DeclareRobustCommand{\textcyr}[1]{\leavevmode{\cyrtext #1}}

\providecommand{\tabularnewline}{\\}

\usepackage{color}

\def\NOT(#1,#2){\OneQubitGate(#1,#2){$X$}}

\makeatother

\begin{document}
\title{Multi-photon multi-quantum transitions in the spin-$\frac{3}{2}$
silicon-vacancy centers of SiC}
\author{Harpreet Singh\textsuperscript{1}, Mario Alex Hollberg\textsuperscript{1},
Andrei N. Anisimov\textsuperscript{2}, Pavel G. Baranov\textsuperscript{2}
and Dieter Suter\textsuperscript{1}\\
 \textsuperscript{1}Fakultät Physik, Technische Universität Dortmund,\\
 D-44221 Dortmund, Germany. \textsuperscript{2}Ioffe Institute, St.
Petersburg 194021, Russia.}
\begin{abstract}
Silicon vacancy centers in silicon carbide are promising candidates
for storing and manipulating quantum information. Implementation of
fast quantum gates is an essential requirement for quantum information
processing. In a low magnetic field, the resonance frequencies of
silicon vacancy spins are in the range of a few MHz, the same order
of magnitude as the Rabi frequencies of typical control fields. As
a consequence, the rotating wave approximation becomes invalid and
nonlinear processes like the absorption and emission of multiple photons
become relevant. This work focuses on multi-photon transitions of
negatively charged silicon vacancies driven by a strong RF field.
We present continuous-wave optically detected magnetic resonance (ODMR)
spectra measured at different RF powers to identify the 1-, 2-, and
3-RF photon transitions of different types of the silicon vacancy
in the 6$H$-SIC polytype. Time-resolved experiments of Rabi oscillations
and free induction decays of these multiple RF photon transitions
were observed for the first time. Apart from zero-field data, we also
present spectra in magnetic fields with different strength and orientation
with respect to the system's symmetry axis.
\end{abstract}
\maketitle

\section{Introduction}

Vacancies in silicon carbide (SiC) are excellent candidates for quantum
sensing and technology applications$\;$\cite{singh2020optical,soltamov-naturecom-19,widmann-nature-14,lukin-prxq-20,baranov-prb-11}
as they can be controlled coherently and have long spin coherence
times $\:$\cite{christle-nature-14,falk-nature-13,kraus-nature-13}.
Apart from that, mature fabrication techniques exist for SiC. Here
we focus on the silicon vacancies in the $6H$-SiC polytype$\;$\cite{singh-prb-21,singh-prb-20,soltamov-prb-21},
where three types of Si vacancies have been identified, labelled $V_{1}$,
$V_{2}$ and $V_{3}$, which correspond to different lattice positions$\;$\cite{sorman-prb-00,baranov-prb-11,biktagirov-prb-18}.
It has been shown that silicon-vacancies at hexagonal sites $h$ corresponds
to $V_{1}$, while $V_{3}$ and $V_{2}$ are at cubic lattice sites
$k_{1}$ and $k_{2}$ $\;$\cite{davidsson-apl-19}. When the vacancies
are negatively charged ($V_{Si}^{-}$) , they have spin 3/2 and the
ground state spin sublevels can be polarized by optical irradiation$\;$\cite{singh-prb-20,soltamov-prb-21}.
In previous work, we measured relaxation rates and optical spin initialization
of $V_{Si}^{-}$ at room temperature$\;$\cite{singh-prb-20,singh-prb-21},
the polarization dependencies of the ZPLs and the ODMR contrast as
a function of temperature$\;$\cite{breev2021inverted}.

$V_{Si}^{-}$ in $6H$-SiC have smaller crystal field splittings than
nitrogen vacancies in diamond. Even at modest radio frequency (RF)
powers, the spin Rabi frequencies are in the MHz range and thus comparable
to the resonance frequency in zero or low magnetic field. Hence, the
RF field drives not only the allowed single-photon transitions, but
we also find absorption of multiple photons, aperiodic evolution of
the system and nonlinear dependencies of effective Rabi frequencies
on the applied RF strength. A multi-photon transition occurs by simultaneous
absorption and emission of multiple photons such that the transition
frequency corresponds to an integer multiple of the driving RF frequency.
Two-photon transitions have been studied in different systems$\;$\cite{nathan1985review,wallis-prb-74,hutchings1992nondegenerate,hayat2008observation}
including NMR of spin-1 system$\;$\cite{pines-prb-78}. The two RF
photon absorption between two spin states differing by $\Delta m_{S}=2$
has also been observed in $V_{Si}^{-}$$\;$\cite{kraus-nature-13}.
Further, the absorption of a single photon causing a weak transition
with $\Delta m_{S}=2$ was also observed in the $V_{2}$ type vacancies
of $4H$-SiC \cite{carter-prb-15}. The authors mentioned as a possible
cause of these transitions stray magnetic fields not aligned with
the $c$-axis or the hyperfine interaction of the $V_{Si}^{-}$ with
nuclear spins which lead to a mixing spin levels and make the double
quantum transitions, which are normally forbidden, weakly allowed.
Later, it has been demonstrated that the amplitude of the double-quantum
transitions of $V_{Si}^{-}$ was the same in natural and isotopically
purified 4H-SiC samples$\;$\cite{simin-prx-16}. These authors considered
additional terms in the spin Hamiltonian due to the trigonal pyramidal
symmetry of this center as the cause of these double-quantum spin
transitions$\;$\cite{simin-prx-16}.

\begin{figure}
\includegraphics[width=0.99\columnwidth]{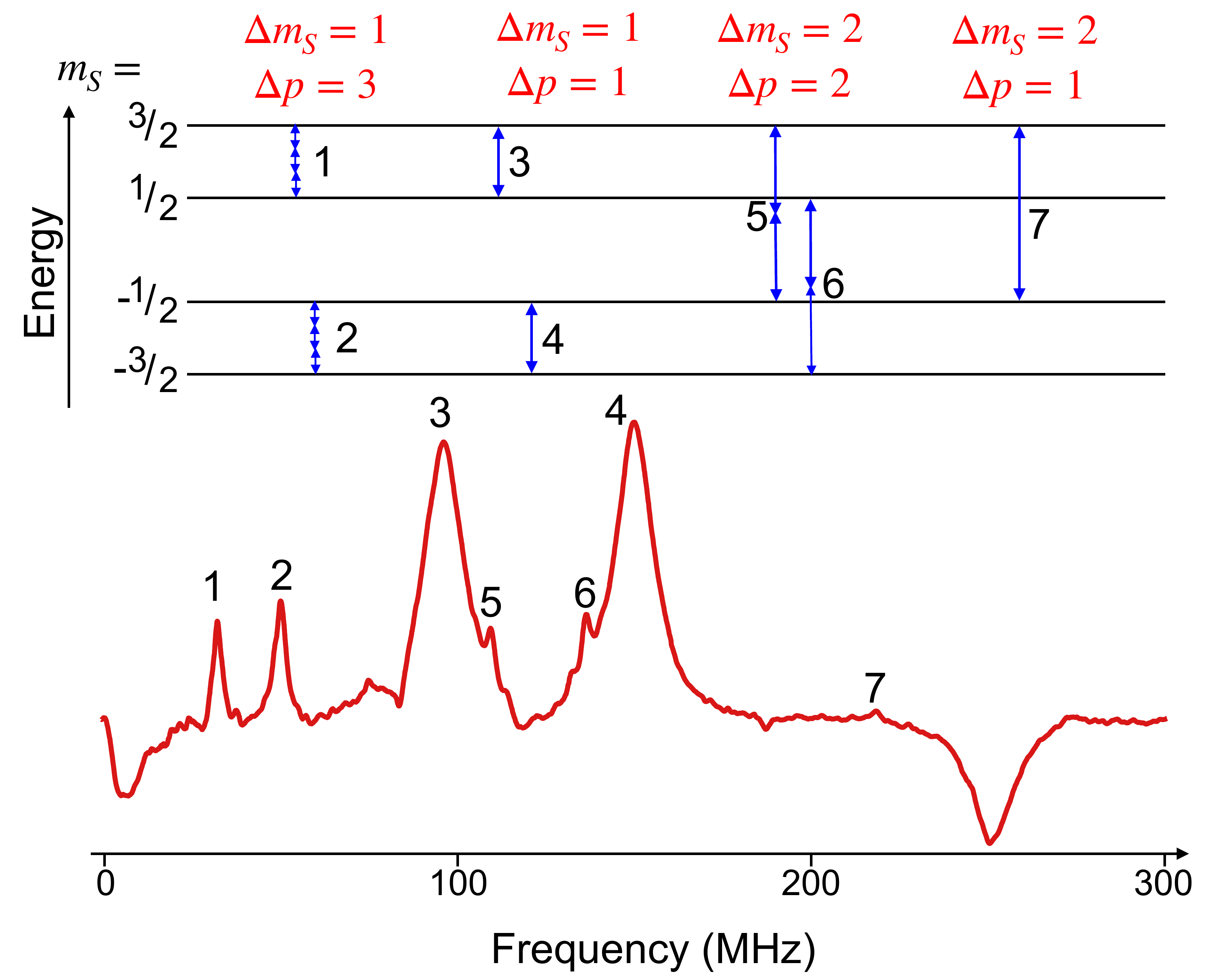}\caption{ODMR spectrum and energy level diagram of the spin states of $V_{3}$
in a 4.5 mT magnetic field $\Vert$ to the c-axis with different types
of multiple photon and multiple quantum transitions. $\Delta m_{s}$
and $\Delta$$p$ are the changes in spin angular momentum and the
number of absorbed RF photons for each transition.}

\label{titlefig}
\end{figure}

Figure$\;$\ref{titlefig} shows a typical experimental spectrum of
the $V_{3}$ vacancy in magnetic field of 4.5 mT $\Vert$ c-axis with
a number of transitions that do not match the ``allowed'' magnetic
dipole transitions. The resonances marked 1 and 2, which are found
at the lowest frequencies, are 3-photon transitions between spin states
with $\Delta m_{s}$=1 while peaks 5 and 6 are due to the 2-photon
transitions between spin levels with $\Delta m_{s}$=2. The peak labelled
with 7 in Fig.$\;\ref{titlefig}$ is due to a single-photon, 2-quantum
transition between the spin levels 3/2 $\leftrightarrow$-1/2.

This work explores the possible multi-photon and multiple quantum
transitions of the $V_{Si}^{-}$ in the $6H$-SiC using continuous
wave and pulsed ODMR techniques. We numerically simulate the stationary
and time-dependent system response through a master equation for the
$V_{Si}^{-}$ spin ensemble. The results are in excellent agreement
with the experimental observations.

This paper is organized as follows. Section$\;$\ref{system} introduces
the properties of the spin system. The first subsection of Sec.$\;$\ref{sec:odmr}
provides some details about continuous-wave optically detected magnetic
resonance (cw-ODMR) experiments at different RF powers. The second
subsection shows time-resolved measurements of 1-, 2-, and 3-RF photon
transitions. Section$\;$\ref{sec:Spin-dynamics} provides some details
about the simulations. Section$\;$\ref{subsec:ODMRmag} contains
the ODMR recorded in the presence of magnetic fields. Finally, section$\;$\ref{sec:conc}
contains a brief discussion and some concluding remarks.

\section{System}

\label{system} 

The experiments reported here were performed on an ensemble of $V_{Si}^{-}$
in a $6H$-SiC sample, and details of the sample preparation are given
in Appendix A. The site symmetry of the $V_{Si}^{-}$ centre is $C_{3v}$
and the electronic spin is $S=$3/2. The spin Hamiltonian is 
\begin{equation}
{\cal H}=D(S_{z}^{2}-\frac{S(S+1)}{3})+g\mu_{B}\vec{B}\cdot\vec{S},\label{eq:hamiltonian}
\end{equation}

where the zero field splitting (ZFS) in the electronic ground state
is $2D$= 128 MHz, in frequency units ($h=1$) for $V_{2}$ and -28
MHz for $V_{3}$$\;$\cite{soltamov-naturecom-19} . $g=2$ is the
electron $g$ factor, $\mu_{B}$ is the Bohr magneton, $\vec{B}$
is the applied static magnetic field and $\vec{S}$ = \{$S_{x},$
$S_{y}$, $S_{z}$\} is the vector of spin operators.

We drive the evolution of the electron spins driven by a sinusoidally
oscillating RF field. For a field perpendicular to the symmetry axis,
the interaction Hamiltonian between the RF field and the spins is

\begin{equation}
{\cal H}_{RF}(t)=\text{\ensuremath{\Omega_{1}}}(\cos\varphi\,cos\,(2\pi\omega t)\:S_{x}+\sin\varphi\,sin\,(2\pi\omega t)\:S_{y}),\label{eq:hrf}
\end{equation}
where $\Omega_{1}=g\mu_{B}B_{1}$ represents the strength of the coupling
to the RF field $B_{1}$ in frequency units, $\omega$ is the oscillation
frequency of the field and the angle $\varphi$ parametrises the polarization
of the RF field: for $\varphi=n\pi/2$, $n$ integer, the field is
linearly polarized, other values correspond to elliptical (including
circular) polarization.

When the $V_{Si}^{-}$ are irradiated with a suitable wavelength laser,
population is transferred to the excited state. Most of the population
falls back to the ground state by spontaneous emission, but some of
it undergoes intersystem crossing (ISC) to the shelving states and
preferentially returns to specific ground-state spin-levels depending
upon the type of vacancies$\;$\cite{baranov-prb-11,soltamov-prl-12,singh-prb-21}:
in $V_{2}$ ($V_{3}$), the population preferentially falls into the
$\pm$3/2 ($\pm$1/2) spin states$\;$\cite{singh-prb-20}.

Another process that affects the dynamics of the silicon-vacancy is
spin relaxation : spin-lattice ($T_{1}$ relaxation) and spin-spin
relaxation ($T_{2}$ relaxation). We studied these relaxation processes
at room temperature and determined the corresponding relaxation rates
in previous works$\;$\cite{singh-prb-20,singh-prb-21}.

\section{Optically detected magnetic resonance}

\label{sec:odmr}

\begin{figure}
\includegraphics[scale=1.1]{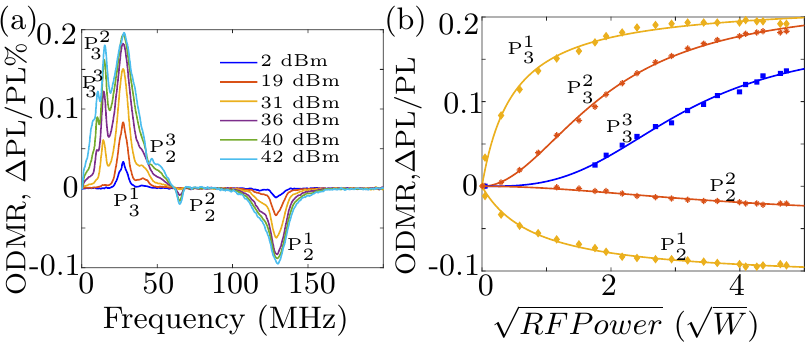}

\caption{a) Experimental ODMR spectra at different power levels. b) Amplitudes
of the different peaks vs. RF field strength.}

\label{cwodmr}
\end{figure}

Optically detected magnetic resonance (ODMR) is a technique for measuring
electron spin resonance (ESR) through an optical signal instead of
inductive detection$\;$\cite{carbonera2009optically,mr-1-115-2020}.
The optical irradiation establishes a non-thermal population of the
different spin states. Since the different spin states contribute
differently to the photoluminescence (PL) rate, a change in the spin
polarization leads to a change of the PL rate, which can be used to
measure the spin polarization. The sensitivity of this technique is
often high enough to measure ODMR of individual electron spins$\;$\cite{SUTER201750,bathen,widmann-nature-14,mr-1-115-2020}.
We used the same technique for measuring the ODMR of silicon vacancies
in 6H-SiC with natural isotopic composition. Additional details of
the ODMR setup are given the Appendix$\;$B.

\begin{table}
\begin{centering}
\begin{tabular}{|c|c|c|c|c|}
\hline 
$V_{Si}^{-}$ & Peaks & $c$ & $S_{max}$(\%) & $\Lambda_{0}$($W^{c/2}$)\tabularnewline
\hline 
\hline 
$V_{2}$ & P$_{2}^{1}$ (128 MHz) & 1 & -0.111$\pm$0.004 & 0.825$\pm$0.117\tabularnewline
\hline 
$V_{2}$ & P$_{2}^{2}$ (64 MHz) & 2 & -0.037$\pm$0.007 & 14.92$\pm$5.34\tabularnewline
\hline 
$V_{3}$ & P$_{3}^{1}$ (28 MHz) & 1 & 0.219$\pm$0.006 & 0.488$\pm$0.076\tabularnewline
\hline 
$V_{3}$ & P$_{3}^{2}$ (14 MHz) & 2 & 0.218$\pm$0.004 & 3.693$\pm$0.236\tabularnewline
\hline 
$V_{3}$ & P$_{3}^{3}$ (9 MHz) & 3 & 0.171$\pm$0.008 & 28.92$\pm$3.59\tabularnewline
\hline 
\end{tabular}
\par\end{centering}
\caption{Fitting parameters of Eq$\;\eqref{eq:powerrf}$ for the different
ODMR peaks P$_{2}^{i=1,2}$ and P$_{3}^{i=1,2,3}$ of the $V_{2}$
and $V_{3}$ vacancies.}

\centering{}\label{fittingparameter}
\end{table}

\subsection{Continuous-wave ODMR}

\label{subsec:Continuous-wave-ODMR}

In our continuous-wave (CW) ODMR experiments, the sample was continuously
illuminated with 785 nm laser light while the RF field was modulated
(switched ON / OFF). The PL from the sample was collected and detected
by an avalanche photodiode (APD) module after passing through a 850
nm long-pass filter. The electrical signal from the APD was recorded
by a lock-in amplifier referenced to the ON/OFF modulation signal$\;$\cite{singh-prb-20,singh-prb-21}.
Figures$\;$\ref{titlefig} and \ref{cwodmr}(a) show two such spectra,
which were recorded at room temperature, in a field of 4.5 mT (Figure$\;$\ref{titlefig})
and 0 (Figure \ref{cwodmr}(a)), respectively, by recording the lock-in
signal as a function of the RF frequency. At low RF power, two peaks
are visible in the zero-field spectrum of Figure$\;$\ref{cwodmr}(a),
a positive one at 28 MHz (P$_{1}^{3}$) and a negative one at 128
MHz (P$_{1}^{2}$), which correspond to the $V_{3}$ and $V_{2}$
sites of $V_{Si}^{-}$. A positive signal corresponds to an increase
in the PL when the RF field is resonant. As the RF power is increased,
the amplitude and width of these peaks increases and additional peaks
become visible at 9 MHz (P$_{3}^{3}$), 14 MHz (P$_{2}^{3}$), 43
MHz (P$_{3}^{2}$), and 64 MHz (P$_{2}^{2}$). The resonance lines
marked P$_{3}^{3}$ (P$_{3}^{2}$) and P$_{2}^{3}$ (P$_{2}^{2}$)
appear at one-third and one-half to the frequency of the $V_{3}$
($V_{2}$) peaks. The peak P$_{3}^{2}$ at 43 MHz overlaps with P$_{1}^{3}$.

The different peaks show remarkably different power dependence as
shown in Figure$\;$\ref{cwodmr}(b), which represents the ODMR signal
amplitude vs. the square root of the applied RF power of all peaks
except P$_{3}^{2}$ (which is difficult to extract, due to the overlap).
The experimental data were fitted with the function
\begin{equation}
S(\Lambda)=S_{max}[\Lambda^{c}/(\Lambda_{0}+\Lambda^{c})],\label{eq:powerrf}
\end{equation}
where $S(\Lambda)$ is the ODMR signal amplitude and $\Lambda$ the
square root of RF power. The exponent $c$ is set to 1, 2 and 3 for
the 1-, 2- and 3-photon transitions. If the exponent $c$ is used
as an additional fitting parameter, the optimal fit is obtained with
values that are close to these. The values for $S_{max}$ and $\Lambda_{0}$
for the different peaks are given in Table$\;$\ref{fittingparameter}.

\subsection{Time-resolved experiments}

For measuring Rabi oscillations and free induction decays with 1-,
2-, and 3 RF photon excitation, we used time-resolved ODMR. A typical
time-resolved ODMR experiment consists of an initial laser pulse,
which polarizes the spin ensemble. Then, a set of RF pulses and delays
are applied before measuring the PL with a measuring laser pulse.
We will start with the Rabi oscillation.

\subsubsection{Rabi oscillations}

\label{subsec:Rabi-Oscillations}

\begin{figure}
\includegraphics{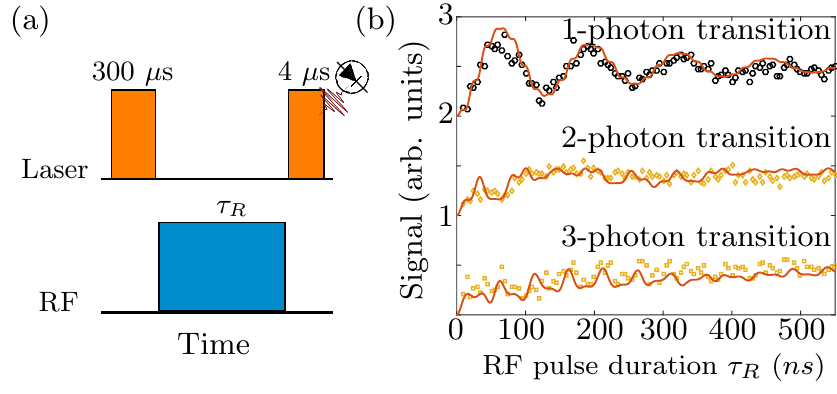}\caption{(a) Pulse sequence for measuring Rabi oscillations . (b) Rabi oscillations
measured for $V_{3}$ with 1-, 2- and 3 RF photons at room temperature.
Circles, diamonds and squares represent experimentally measured data.
Red curves are the simulations obtained by solving Eq.$\;$\ref{eq:lindblad}
numerically for different photons transitions. The curves have been
shifted vertically to avoid overlap.}

\label{rabi}
\end{figure}

Figure$\;$\ref{rabi} (a) shows the pulse sequence for measuring
Rabi oscillations for the different transitions. The spin ensemble
was first polarized with a 300 $\mu$s laser pulse. A signal was recorded
during a 4 $\mu$s laser pulse after an RF pulse of variable duration
$\tau_{R}$. The same experiment was repeated without the RF pulse
to obtain a reference signal to suppress background signals. The signals
obtained from both experiments were subtracted from each other for
each pulse duration $\tau_{R}$$\;$\cite{singh-prb-20,singh-prb-21}.
Figure$\;$\ref{rabi} (b) shows the Rabi oscillation recorded in
zero magnetic field for the $V_{3}$ type with 1-, 2- and 3-RF photon
transitions at room temperature. For measuring Rabi oscillations of
the 1 photon transition, a 28 MHz linearly polarized RF pulse of 11
W was applied, whereas 14 MHz (9.3 MHz) was used for the 2 (3) photon
transition with the same RF power. The red curves in Fig.$\;$ \ref{rabi}
(b) are signals calculated for the 1 photon ($\omega$=28 MHz), 2
photon ($\omega$=14 MHz), and 3 photon ($\omega$=9.3 MHz) transitions.
More details about the simulations of the curves are given in subsection$\;$\ref{subsec:Evolution-under-Hamiltonian}.

\begin{figure}
\includegraphics{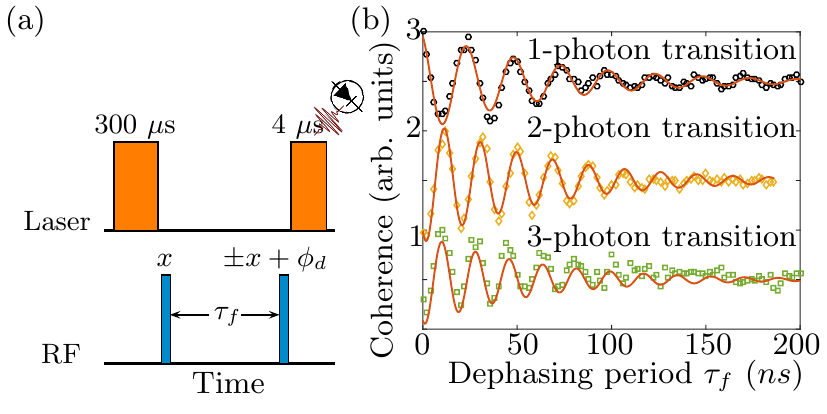}\caption{(a) Pulse sequence for measuring FIDs . (b) FIDs measured for $V_{3}$
with 1-, 2- and 3 RF photons in zero magnetic field at room temperature.
Circles, diamonds and squares represent experimentally measured data.
Red curves were obtained by fitting the experimental data to the fitting
function given in Eq.$\:$\eqref{eq:fidfit}. The curves have been
shifted vertically to avoid overlap.}

\label{fid0B}
\end{figure}

\subsubsection{Free Induction decay}

Figure$\;$\ref{fid0B} (a) shows the pulse sequence for measuring
free induction decays (FIDs). As above, we initialized the spin ensemble
by a first laser pulse and a first RF pulse with a frequency of 28
MHz, 14 MHz, and 9.3 MHz for 1-, 2-, and 3 photon transitions and
duration 25 ns for all cases . During the subsequent delay $\tau_{f}$,
the coherence generated by the pulse was allowed to precess freely.
A second RF pulse with the same frequency and duration, and phase
$\phi_{d}=2\pi f_{det}\tau_{f}$ (detuning frequency $f_{det}$=-40
MHz) was applied before the readout laser pulse. We again used the
difference between two experiments, where the second RF pulses have
phases $\phi_{d}$ and $\phi_{d}+\pi$, respectively, to suppress
unwanted background signals$\;$\cite{singh-prb-20,singh-prb-21}.
Figure$\;$\ref{fid0B} (b) shows the experimentally recorded FIDs
with the 1-, 2-, and 3 RF photon pulses in zero magnetic field and
at room temperature. The recorded experimental data were fitted with
the function:

\begin{eqnarray}
S_{x+\phi_{d}}^{N}-S_{-x+\phi_{d}}^{N} & = & A\:cos(2\pi f-\phi)\:e^{-\tau_{f}/T_{2}^{*}},\label{eq:fidfit}
\end{eqnarray}

where $S_{x+\phi_{d}}^{N}$ and $S_{-x+\phi_{d}}^{N}$ are signals
recorded with $N$-photon pulses in the main and reference experiment.
The values obtained for the frequency $f$ are 40.0$\pm$0.1 MHz,
53.3$\pm$0.2 MHz, and 56.1$\pm$0.5 MHz for 1-, 2-, and 3 photon
FID measurements respectively; and the average dephasing rate $T_{2}^{*}$
is 62.0$\pm$ 5.6 $ns$. In these experiments, the signal frequency
$f$ is the difference between the transition frequency ($2D$ = 28
MHz) and the sum of the RF frequency $f_{RF}$ and the detuning frequency
$f_{det}$ : $f=|2D-(f_{RF}+f_{det})|$.

\section{Spin dynamics}

\label{sec:Spin-dynamics}

The dynamics of the closed spin system can be estimated using the
Liouville--von Neumann equation:

\begin{eqnarray}
\frac{\partial\rho}{\partial t} & = & -2\pi i\:[{\cal H}_{t}(t),\rho],\label{eq:neumanneqn}
\end{eqnarray}

where we use units with $h=1$, $\rho$ is the spin density matrix,
${\cal H}_{t}(t)={\cal H}+{\cal H}_{RF}(t)$ is the total Hamiltonian,
i.e., the sum of the static system Hamiltonian ${\cal H}$ given in
Eq.$\;$\eqref{eq:hamiltonian} and the time-dependent RF Hamiltonian
${\cal H}_{RF}(t)$ in the laboratory frame which is given in Eq.$\;$\eqref{eq:hrf}.
We solve it numerically using the Runge-Kutta 4 method, without invoking
the rotating wave approximation to include higher-order contributions.

\subsection{Coherent Evolution}

\label{subsec:Evolution-under-Hamiltonian}

\begin{figure}
\includegraphics{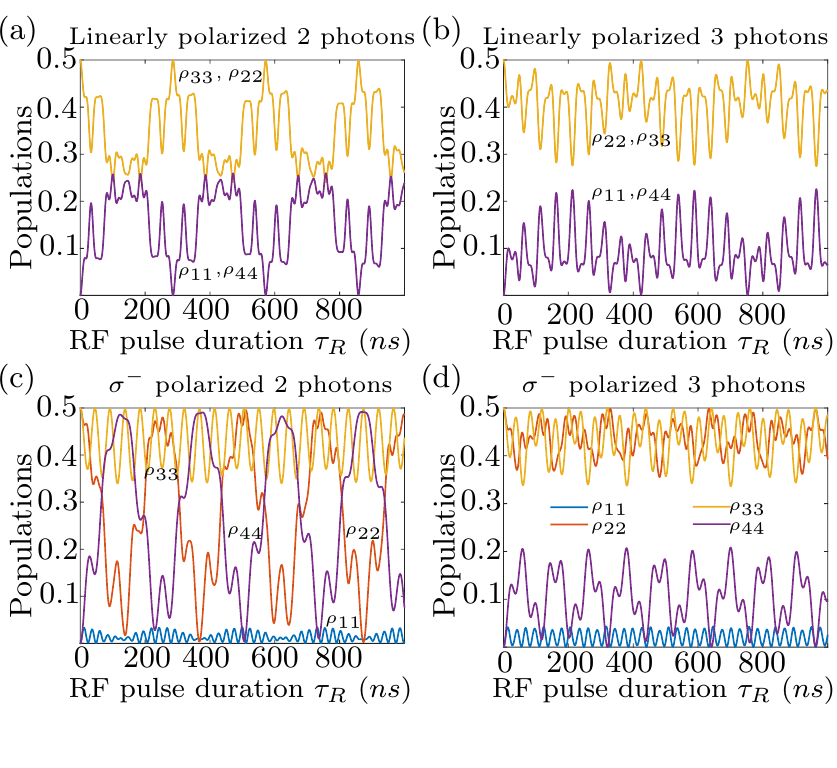}\caption{Simulation of population dynamics of the $V_{3}$ vacancy with linearly
polarized RF excitation for (a) 2 photons (14 MHz), (b) 3 photons
(9.3 MHz), left circularly ($\sigma^{-}$) polarized (c) 2 photons
(14 MHz) , and (d) 3 photons (9.3 MHz).}

\label{simpop}
\end{figure}

Figure$\;$\ref{simpop} shows the simulation results obtained using
equation Eq$\;$\eqref{eq:neumanneqn} (ignoring relaxation processes),
with a diagonal initial state $\rho_{0}=\{0,0.5,0.5,0\}$, RF coupling
strength $\Omega_{1}$ = 9.1 MHz, using the RF Hamiltonian given in
Eq$\;$\eqref{eq:hrf} with $\phi$=$0$ for linear and $\phi$=$\pi/4$
for left circular ($\sigma^{-}$) polarized RF. Figure$\;$\ref{simpop}$\;$(a)
and (b) shows the population dynamics of $V_{3}$ with the absorption
of 2 ($\omega$=14 MHz) and 3 ($\omega$=9.3 MHz) linearly polarized
RF photons, respectively. In comparison, Figure$\;$\ref{simpop}$\;$(c)
and (d) shows the population dynamics for the system interacting with
a left circularly ($\sigma^{-}$) polarized RF field resonant with
the 2- and 3- photon transitions. The circularly polarised field breaks
the symmetry observed in Fig.$\;$\ref{simpop}$\;$(a) and (b) by
exciting mostly the spin transition 1/2$\leftrightarrow$-3/2 with
$\Delta m_{S}=2$ in the case of 2-photon absorption and the transitions
1/2$\leftrightarrow$-3/2 and $-$1/2$\leftrightarrow$-3/2 with $\Delta m_{S}=2$
and $\Delta m_{S}=1$ in the case of 3-photon excitation. This is
in contrast to the linearly polarized field which couples equally
to 1/2$\leftrightarrow$-3/2 and -1/2$\leftrightarrow$3/2 and to
1/2$\leftrightarrow$-3/2 and -1/2$\leftrightarrow$3/2.

\begin{figure}
\includegraphics{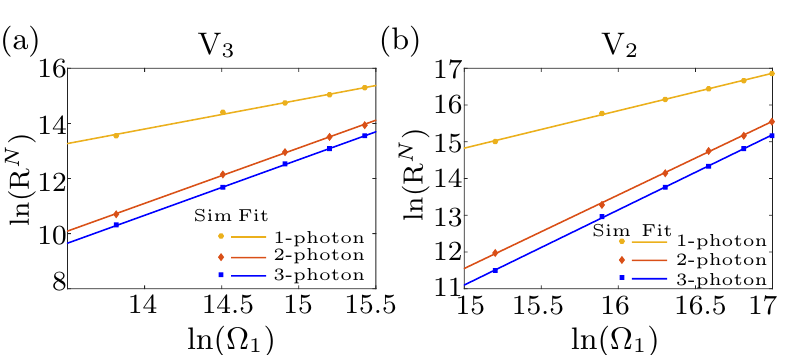} \caption{Log of Rabi frequency of $N$-photon transitions vs. Log of RF coupling
strength $\Omega_{1}$ for (a) $V_{3}$ and (b) $V_{2}$.}

\label{rabivsomrga1}
\end{figure}

Since the strength of the RF field is comparable to the transition
frequency, the relation between field strength and the observed dynamics
is highly nonlinear. As shown in Figure$\;$\ref{simpop}, the oscillations
of the populations contain multiple frequencies. Fourier-transformation
shows that they consist of a small number of frequencies whose amplitudes
and positions shift with increasing RF field strength. As a comparison
with typical resonant excitations (Rabi-flopping), we perform Fourier
transforms of the time-traces and consider the main frequency component
as the Rabi frequency. For the data of Figure$\;$\ref{simpop}, the
extracted Rabi frequencies are 7.8 MHz, 3.39 MHz, and 2.56 MHz for
1-, 2-, 3-photon transitions. In Fig.$\;$\ref{rabivsomrga1}, we
show additional data for different RF field strengths in log-log plots.
The individual data points are fitted to straight lines, 
\begin{eqnarray}
ln\;R_{N} & = & a\;ln\;\Omega_{1}+b.\label{eq:log-log}
\end{eqnarray}
The fitting parameters $a$ and \textbf{$b$} are given in Table$\;$\ref{loglogtable}.

\begin{table}
\begin{centering}
\begin{tabular}{|>{\centering}p{1.5cm}|c|c|c|c|}
\hline 
\multicolumn{5}{|c|}{Log-log fitting}\tabularnewline
\hline 
\hline 
 & \multicolumn{2}{c|}{$V_{3}$} & \multicolumn{2}{c|}{$V_{2}$}\tabularnewline
\hline 
$N$ & $a$ & $b$ & $a$ & $b$\tabularnewline
\hline 
1 & 1.05$\pm$0.10 & -0.95$\pm$1.55 & 1.02$\pm$0.03 & -0.47$\pm$0.51\tabularnewline
\hline 
2 & 2.01$\pm$0.05 & -17.07$\pm$0.77 & 2.00$\pm$0.08 & -18.51$\pm$1.35\tabularnewline
\hline 
3 & 2.02$\pm$0.02 & -3.52$\pm$0.30 & 2.05$\pm$0.02 & -19.59$\pm$0.35\tabularnewline
\hline 
\end{tabular}\caption{Fitting parameters of Eq.\ref{eq:log-log}.}
\par\end{centering}
\centering{}\label{loglogtable}
\end{table}

\subsection{Relaxation}

\begin{figure}
\includegraphics{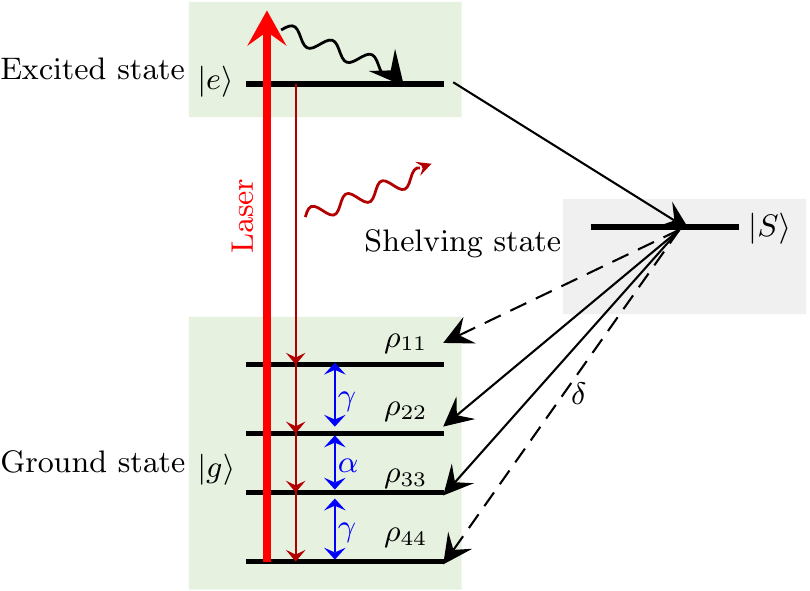}

\caption{Simplified energy-level diagram of $V_{Si}^{-}$, indicating optical
pumping and relaxation processes. The ground, excited, and shelving
states are labeled $\vert g\rangle$, $\vert e\rangle,$ and $\vert s\rangle$.
The nonresonant laser excitation is shown with a red arrow.}

\label{opticalpumping}
\end{figure}

\label{subsec:Relaxation}

The $V_{Si}^{-}$ spin ensemble continuously interacts with its environment,
which causes relaxation. This system-environment interaction has two
main effects on the system, which is not in its thermal equilibrium
state. The first is the loss of coherence, also called spin-spin relaxation,
which preserves the energy of the system. It can be taken into account
by a term $\dot{\rho}_{ik}=-\rho_{ik}/T_{2}^{ik}$ in the equation
of motion. The second effect is that the populations relax back to
thermal equilibrium by a process called spin-lattice relaxation. In
this process, energy is exchanged between the system and its environment
(the lattice)$\;$\cite{abragam-book}. This contribution can be taken
into account as $\dot{\vec{p}}=M\vec{p}$, where the population vector
$\vec{p}$ includes the diagonal density operator elements and the
transition matrix $M$ depends on the mechanism that couples the system
to the environment. Here, we assume that the spin levels 1/2 and -1/2
equilibrate with a rate $\alpha$ and spin levels $\pm3/2$ and $\pm$1/2
equilibrate with a rate $\gamma$. This corresponds to the transition
matrix 
\begin{eqnarray*}
M & = & \left(\begin{array}{cccc}
-\gamma & \gamma\\
\gamma & -\alpha-\gamma & \alpha\\
 & \alpha & -\alpha-\gamma & \gamma\\
 &  & \gamma & -\gamma
\end{array}\right).
\end{eqnarray*}
In our previous work \cite{singh-prb-21}, we determined these spin
relaxation rates for the $V_{1}$ type of $V_{Si}^{-}$ and found
that the ratio $\alpha/\gamma$ agrees with the theoretical value
of 4/3. We also studied the dynamics of the optical initialization
process and found that the laser illumination transfers the population
from $\pm3/2$ to $\pm$1/2 with a pumping rate $\delta$, which is
proportional to the laser intensity$\;$\cite{singh-prb-21}. Figure$\;$\ref{opticalpumping}
shows the optical pumping and relaxation schemes used for the silicon
vacancies. The ground, excited, and shelving states are labeled $\vert g\rangle$,
$\vert e\rangle$, and $\vert s\rangle$. The nonresonant laser excitation
is shown with a thick red arrow. When the vacancy is excited with
the laser light, most of the population falls back from the electronically
excited state to the ground state by spontaneously emitting photons.
The system can also undergo ISC to the shelving states, from where
it preferentially to the spin levels $\pm1/2$ ( $\pm3/2$ ) of the
electronic ground state in case $V_{3}$ ($V_{2}$). An optical pumping
rate $\delta$ is the rate at which population is pumped from the
electronic ground state spin levels $\pm3/2$ to $\pm1/2$ in case
of $V_{3}$ and vice versa for $V_{2}$$\;$\cite{singh-prb-21,singh-prb-20}.

For the numerical simulation, we use the Lindblad master equation
\begin{eqnarray}
\frac{\partial\rho}{\partial t} & = & -2\pi i\:[{\cal H}_{t}(t),\rho]+\sum_{\alpha,\beta,\delta_{1,..,5}}L_{i}^{\dagger}.\rho.L_{i}-\frac{1}{2}\left\{ L_{i}^{\dagger}L_{i},\rho\right\} \label{eq:lindblad}
\end{eqnarray}
here $L_{\alpha}=\sqrt{2}\left(\begin{array}{cccc}
0 & \sqrt{\gamma} & 0 & 0\\
\sqrt{\gamma} & 0 & \sqrt{\alpha} & 0\\
0 & \sqrt{\alpha} & 0 & \sqrt{\gamma}\\
0 & 0 & \sqrt{\gamma} & 0
\end{array}\right)\thickapprox\sqrt{2\alpha}\:S_{x}$ (taking into account $\gamma=3\alpha/4$) drives the spin-lattice
relaxation process, $L_{\beta}=\sqrt{2\beta}\:S_{z}$ is the Lindblad
operator for the dephasing process, and $L_{\delta_{1,..,5}}$ are
the Lindblad operators for the optical pumping . The matrix forms
of the other Lindblad operators are given in APPENDIX C.

The red curves in Fig. 4 (b) of subsection$\;$\ref{subsec:Rabi-Oscillations}
are signals calculated as 0.5(1+$\rho_{11}$-$\rho_{22}$-$\rho_{33}$+$\rho_{44}$)
from Eq$\;$\eqref{eq:lindblad} for 1 photon- ($\omega$=28 MHz),
2 photon- ($\omega$=14 MHz), and 3 photon- ($\omega$=9.3 MHz) transitions,
with an initial diagonal state $\rho_{0}=\{0,0.5,0.5,0\}$, RF coupling
strength $\Omega_{1}=9.1$ MHz, using the RF Hamiltonian given in
Eq$\;$\eqref{eq:hrf} with $\phi=0$ for a linearly polarized RF
field and using the Runge-Kutta 4 method. Additional details are given
in Appendix C. To simulate the effect of RF inhomogeneity, signals
with 0.97, 1 and 1.03 times $\Omega_{1}$, were calculated and added
with weights 0.3, 0.4 and 0.3, respectively. The values of the relaxation
rate was $\alpha=1/107\;\mu s^{-1}$ and for the dephasing rate $\beta$
= 1/148 ns$^{-1}$ , 1/500 ns$^{-1}$, and 1/1000 ns$^{-1}$ for the
1-, 2-, and 3-photon excitation.

\begin{figure}
\includegraphics{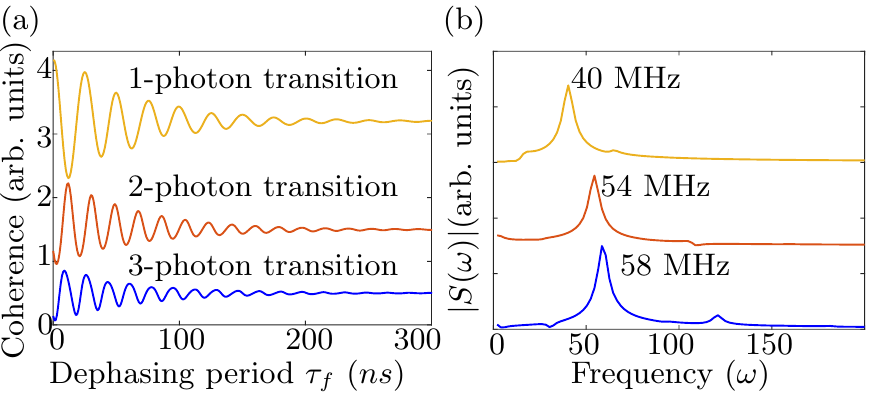}

\caption{(a) Simulated FIDs for 1-, 2-, and 3 photon excitation using Eq$\;$\eqref{eq:lindblad},
and (b) Fourier transform of the simulated FID signals.}

\label{simfid}
\end{figure}

Figure$\;$\ref{simfid} shows the simulated FIDs for the 1-, 2-,
and 3 RF photon excitation using equation Eq$\;$\eqref{eq:lindblad}
with a diagonal, spin-lattice relaxation rate $\alpha$=1/107 $\mu$s$^{-1}$.
For the FID simulations, we used an initial state $\rho_{0}=\{0,0.5,0.5,0\}$
to which we applied a 25 ns RF pulse by letting the system evolve
under the total Hamiltonian ${\cal H}_{t}(t)={\cal H}+{\cal H}_{RF}(t)$,
with the RF coupling strength $\Omega_{1}$ = 9.1 MHz. The resulting
state was allowed to evolve for a time $\tau_{f}$ under the system
Hamiltonian ${\cal H}$, with the dephasing rate $\beta$=1/62 ns$^{-1}$.
After the free evolution, the second RF pulse was applied, with the
phase $\phi_{d}=2\pi f_{d}\tau_{f}$ ($f_{d}=-40$ MHz). The resulting
signal was calculated from the final density matrix elements as $S_{\ensuremath{x+\phi_{d}}}^{N}=\rho_{11}-\rho_{22}-\rho_{33}+\rho_{44}$.
Similarly the reference signal $S_{-x+\phi_{d}}^{N}$was calculated
by adding an extra $\pi$ phase in the RF Hamiltonian of the second
pulse. The difference signal $S_{\ensuremath{x+\phi_{d}}}^{N}-S_{-x+\phi_{d}}^{N}$
is plotted in Figure$\;$\ref{simfid} (a) while Figure$\;$\ref{simfid}
(b) shows the corresponding Fourier transforms (FFTs) for the 1-,
2-, and 3 RF photon excitation. The main resonance lines are found
at $f=|2D-(f_{RF}+f_{det})|$, in agreement with the experimental
data.

\subsection{CW ODMR}

\label{subsec:CW-ODMR}

The data in subsection$\;$\ref{subsec:Continuous-wave-ODMR} clearly
show that 1-, 2-, and 3-photon transitions show different dependence
on the RF power. It is therefore essential to model the system dynamics
under different RF power levels and in constant laser light. To obtain
the CW ODMR spectra, we integrated the master equation with the total
Hamiltonian ${\cal H}_{t}(t)={\cal H}+{\cal H}_{RF}(t)$ for 4 $\mu$s
and plotted the resulting signal as a function of the RF frequency.
The ODMR signal was calculated as the difference $S_{RF}-S_{0}$ between
the signal $S_{RF}$ with applied RF field and the reference signal
$S_{0}$. Each part was calculated from the diagonal elements of the
stationary density matrix as $S_{RF}=$ $\rho_{11}-\rho_{22}-\rho_{33}+\rho_{44}$$\;$\cite{nagy2019high,singh-prb-21},
where $\rho_{11}$, $\rho_{44}$ and $\rho_{22}$, $\rho_{33}$ are
the populations of the spin levels$\vert\pm3/2\rangle$ and $\vert\pm1/2\rangle$,
respectively.

\begin{figure}
\includegraphics{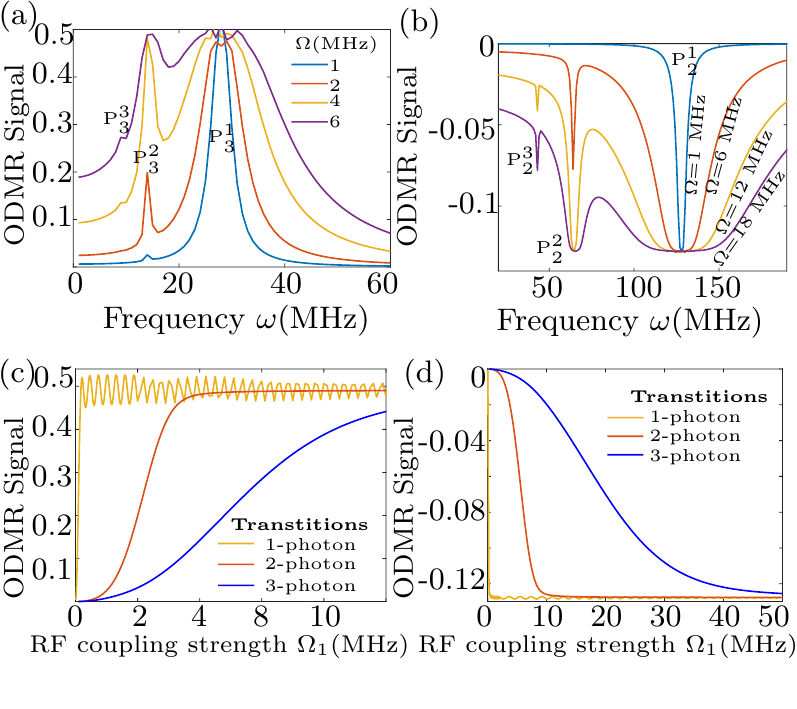}

\caption{Simulated ODMR signal vs frequency $\omega$ at different RF coupling
strength $\Omega_{1}$ for (a) $V_{3}$, (b) $V_{2}$. ODMR signal
of 1-, 2- and 3-photon peaks vs RF coupling strength for (c) $V_{3}$,
(d) $V_{2}$.}

\label{simodmr}
\end{figure}

Figure$\;$\ref{simodmr} (a) shows the simulated ODMR signal at different
RF coupling strengths $\Omega_{1}$ vs RF frequency $\omega$ obtained
by numerically solving Eq.$\;$\eqref{eq:lindblad} with a diagonal
initial state $\rho_{0}=\{\frac{\gamma}{4(\gamma+\delta)},\frac{\gamma+2\delta}{4(\gamma+\delta)},\frac{\gamma+2\delta}{4(\gamma+\delta)},\frac{\gamma}{4(\gamma+\delta)}\}$
($\{\frac{\gamma+2\delta}{4(\gamma+\delta)},\frac{\gamma}{4(\gamma+\delta)},\frac{\gamma}{4(\gamma+\delta)},\frac{\gamma+2\delta}{4(\gamma+\delta)}\}$
for $V_{2}$) where $\gamma=\frac{3}{4}\alpha$, for time $t=$ 0
to 4 $\mu$s in 8000 time steps, using the RF Hamiltonian given in
Eq$\;$\eqref{eq:hrf} with $\phi=0$ for a linearly polarized RF
field. The rates $\alpha$, $\beta$, and $\delta$ were extracted
from experimental data: the spin-lattice relaxation rate is $\alpha$
= 10.8 ms$^{-1}$ (9.3 ms$^{-1}$) for $V_{2}$($V_{3}$), as determined
in our earlier works \cite{singh-prb-21,singh-prb-20}. Under the
influence of an RF field, the dephasing is slower than during free
precession and depends on the homogeneity of the RF field. We therefore
estimated it from the Rabi measurements at different RF powers. For
low power, the dephasing rate decreases linearly with increasing RF
power and then increases linearly in both types of vacancies. This
indicates that the RF field decouples the vacancy spin from nuclear
spins coupled by hyperfine interactions, while at higher power, the
effect of RF inhomogeneity starts to dominate. Further, the dephasing
rate also depends on the applied static magnetic field to the $V_{Si}^{-}$$\;$\cite{carter-prb-15}.
For our calculation, we took the smallest observed dephasing rate
$\beta$$\approx$ 2.5 $\mu$s$^{-1}$(1.3 $\mu$s$^{-1}$) for $V_{2}$($V_{3}$)
to optimise the resolution. The pumping rate $\delta$=1.4 ms$^{-1}$(6.8
ms$^{-1}$) for $V_{2}$($V_{3}$) was extracted from the time-dependence
of the PL signal measured after an RF pulse of 20 $\mu$s using the
model described in Ref.$\;$\cite{singh-prb-21}. Figure$\;$\ref{simodmr}
(b) shows the ODMR signal for the 1-, 2- and 3 RF photon peaks vs.
the RF field strength $\Omega_{1}$. In the simulated ODMR spectra
of the V2 and V3 vacancies, the 1-, 2-, and 3- RF photons absorption
peaks are visible. The intensity of peak $P_{3}^{3}$ (3 RF photon
transition) is small relative other two peaks as compared to the experimental
recorded spectra and we will address this in Sec.$\;$\ref{subsec:RF-field-parallel}.

\section{ODMR in a magnetic field}

While the experiments described in the preceding sections were performed
in zero field, we now consider the effect of magnetic fields. In subsection
\ref{subsec:ODMRmag}, we consider fields applied parallel to the
symmetry axis of the center lifts the degeneracy of the energy levels
as well of the transitions. This is particularly relevant as it allows
one to distinguish between single- (e.g. +1/2 $\leftrightarrow$ +3/2)
and multiple quantum transitions (e.g. -1/2 $\leftrightarrow$ +3/2)
that are degenerate in zero field but behave differently. In subsection
\ref{subsec:Magnetic-field-with}, we also consider fields not aligned
with the symmetry axis, which modify not only the transition frequencies
but also the transition probabilities.

\begin{figure}
\includegraphics{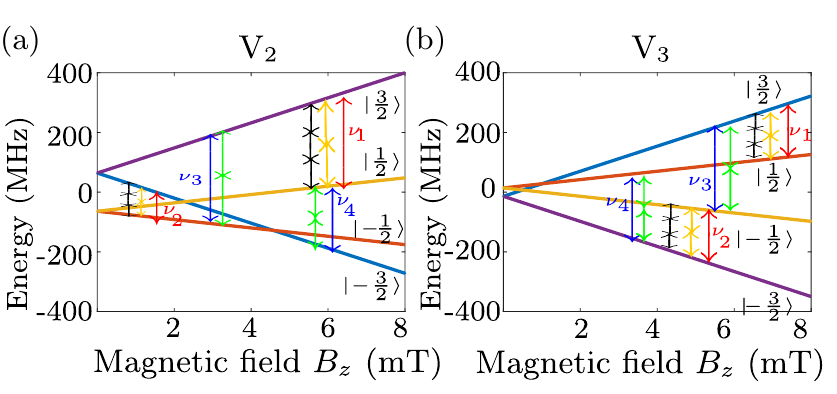}

\caption{Energy levels of (a) the $V_{2}$ vacancy and (b) $V_{3}$ vacancy
in a magnetic field parallel to crystal $c$-axis. Arrows represent
the possible transitions. Red and blue arrows are for the one-photon
transitions; green and yellow arrows are for the two-photon transitions;
and black arrows are for the three-photon transitions.}

\label{energylevels}
\end{figure}

Figure$\;$\ref{energylevels}$\;$(a) and (b) shows the energy levels
of $V_{2}$ and $V_{3}$ types of vacancies in an external magnetic
field parallel to the crystal $c$-axis along with different types
(multiple photon / multiple quantum observe in the experiment) transitions.
The red arrows labeled with $\nu_{1}$ and $\nu_{2}$ represent the
allowed single-quantum transition from $\vert3/2\rangle\leftrightarrow\vert1/2\rangle$
and $\vert-3/2\rangle\leftrightarrow\vert-1/2\rangle$, respectively.
The blue arrows labeled with $\nu_{3}$ and $\nu_{4}$ represent the
two-quantum transitions from $\vert3/2\rangle\leftrightarrow\vert-1/2\rangle$
and $\vert-3/2\rangle\leftrightarrow\vert1/2\rangle$, respectively.

\begin{figure}
\includegraphics{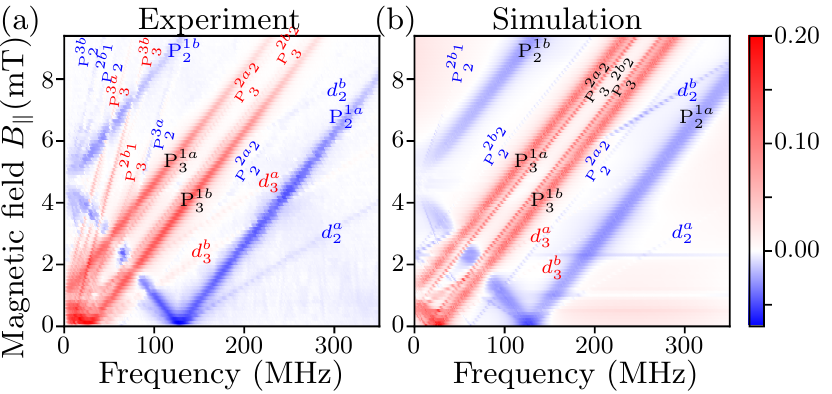} \caption{(a) Experimental, and (b) simulated ODMR spectra for a range of magnetic
fields $B\parallel c$ axis showing different possible transitions
for $V_{1}/V_{3}$ and $V_{2}$. Peaks labeled with $P_{2}^{1a(b)}$
($P_{3}^{1a(b)}$), $P_{2}^{2a(b)}$ ($P_{3}^{2a(b)}$ ) and $P_{2}^{3a(b)}$
($P_{3}^{3a(b)}$ ) are 1, 2 and 3, RF photons absorption peaks due
to the single quantum spin transition $\Delta$$m_{s}=1$, peaks $d_{2}^{a(b)}$($d_{3}^{a(b)}$)
are 1 RF photon absorption peaks due to double quantum spin transition
$\Delta$$m_{s}=2$ for $V_{2}$ ( $V_{3}$ ) type of $V_{Si}^{-}$.
The color scale is in units of $\Delta$PL/PL\%.}

\label{magfield}
\end{figure}

In APPENDIX C, Fig.$\;$\ref{trasitionfromeig} shows different possible
transitions for the $V_{1}/V_{3}$ and $V_{2}$ for a range of magnetic
fields B $\parallel$ c axis, calculated using eigenvalues of Hamiltonian
given in Eq.$\;$\eqref{eq:hamiltonian}.

\subsection{Magnetic field parallel to the $c$-axis}

\label{subsec:ODMRmag}

\begin{table}
\begin{centering}
\begin{tabular}{|c|c|c|c|c|}
\hline 
Peak & $\Delta p$ & $\Delta m_{s}$ & Slope $B_{\Vert}/\omega$ & Transition\tabularnewline
\hline 
\hline 
$P_{i=2,3}^{1a(b)}$ & 1 & 1 & $m$ & $3/2\leftrightarrow1/2$ ($-3/2\leftrightarrow-1/2)$\tabularnewline
\hline 
$d_{i=2,3}^{a(b)}$ & 1 & 2 & $0.5m$ & $3/2\leftrightarrow-1/2$ ($-3/2\leftrightarrow1/2)$\tabularnewline
\hline 
$P_{i=2,3}^{2a_{1}(b_{1})}$ & 2 & 1 & $2m$ & $3/2\leftrightarrow1/2$ ($-3/2\leftrightarrow-1/2)$\tabularnewline
\hline 
$P_{i=2,3}^{2a_{2}(b_{2})}$ & 2 & 2 & $m$ & $3/2\leftrightarrow-1/2$ ($-3/2\leftrightarrow1/2)$\tabularnewline
\hline 
$P_{i=2,3}^{3a(b)}$ & 3 & 1 & $3m$ & $3/2\leftrightarrow1/2$ ($-3/2\leftrightarrow-1/2)$\tabularnewline
\hline 
\end{tabular}
\par\end{centering}
\caption{Detail of Figure$\;$\ref{magfield} peak labels and their correspondence
to the spin transition levels. $\Delta m_{s}$ and $\Delta$$p$ are
the changes in spin angular momentum and the number of absorbed RF
photons for each transition and $m=(g\mu_{B})^{-1}$.}
\label{peaklabel}
\end{table}

Figure$\;$\ref{magfield}$\;$(a) shows the ODMR recorded in an external
magnetic field parallel to the $c$-axis using an RF power of $\thickapprox$
5 W at room temperature. A peak $P_{i=2,3}^{1a(b)}$ is due to the
1 RF photon transition between the levels $3/2\leftrightarrow1/2$
($-3/2\leftrightarrow-1/2$) of the $V_{i=2,3}$ type of $V_{Si}^{-}$.
Peaks labeled with 3 and 4 in Fig.$\;$\ref{titlefig} are 1 RF photon
transition between $3/2\leftrightarrow1/2$ and $-3/2\leftrightarrow-1/2$
of $V_{3}$. Peaks $P_{i}^{2a_{1}(b_{1})}$ are due to the absorption
of 2 RF photons between the spin levels $3/2\leftrightarrow1/2(-3/2\leftrightarrow-1/2)$
(as shown in Fig$\;$\ref{energylevels}$\;$(a) and (b) with two
yellow arrows) and appear at half the corresponding transition frequency.
Another 2 RF photon peaks $P_{i}^{2a_{2}(b_{2})}$ due to absorption
of 2 photons between the spin levels $3/2\leftrightarrow-1/2(-3/2\leftrightarrow1/2)$
as shown in Fig$\;$\ref{energylevels}$\;$(a) and (b) with two green
arrows. Peaks labeled with 5 and 6 in Fig.$\;$\ref{titlefig} are
2 RF photons transition between $3/2\leftrightarrow-1/2$ and $-3/2\leftrightarrow1/2$
of $V_{3}$. A peak $P_{i}^{3a(b)}$ is due to the absorption of 3
RF photons between the spin levels $3/2\leftrightarrow1/2(-3/2\leftrightarrow-1/2)$
(as shown in Fig$\;$\ref{energylevels}$\;$(a) and (b) with three
black arrows) and appears at one third of the corresponding transition
frequency. Peaks labeled with 1 and 2 in Fig.$\;$\ref{titlefig}
are 3 RF photons transition between $3/2\leftrightarrow1/2$ and $-3/2\leftrightarrow-1/2$
of $V_{3}$. The slope of 2- and 3-photon transition between the levels
$\pm3/2\leftrightarrow\pm1/2$ is 2 and 3 times higher than the slope
of the 1-photon transition. Additional peaks $d_{i=2,3}^{a}$ and
$d_{i=2,3}^{b}$ for $V_{i=2,3}$ type of $V_{Si}^{-}$ are due to
the absorption of 1 RF photon by spin levels $\pm3/2\leftrightarrow\mp1/2$.
This type of transitions were also seen previously for the $V_{2}$
type of vacancies in $4H$-SiC, and the authors suggested that the
cause of these transitions was the fine structure of the $V_{2}$
vacancy$\;$\cite{simin-prx-16,Sosnovsky-prb-21}. Details of these
peaks are given in Table$\;$\ref{peaklabel}. Figure$\;$\ref{magfield}$\;$(b)
shows the corresponding simulated spectra for an external magnetic
field almost parallel (angle of inclination is 2.5°) to the $c$-axis,
obtained by numerically solving Eq.$\;$\ref{eq:lindblad} using the
same method and relaxation and pumping parameters are given in the
above subsection \ref{subsec:Continuous-wave-ODMR}. The applied magnetic
field removes the degeneracies in the spin levels. For some magnetic
field values, level-anticrossing also happens, so the reference signal
$S_{\ensuremath{0}}$ was calculated from the diagonal elements of
the stationary density matrix without RF for a particular magnetic
field. The obtained ODMR signal was renormalized to match the experimental
signal and plotted. The RF Strength $\Omega_{1}$ used for both vacancies
was 6 MHz. In the simulations, 1- and 2- photon peaks are visible
along with the 1 RF photon absorption peaks $d_{i=2,3}^{a(b)}$ by
spin levels $\pm3/2\leftrightarrow\mp1/2$ of $V_{i=2,3}$ type of
$V_{Si}^{-}$.

\subsection{Magnetic field with component perpendicular to the $c$-axis}

\label{subsec:Magnetic-field-with}

\begin{figure}
\includegraphics{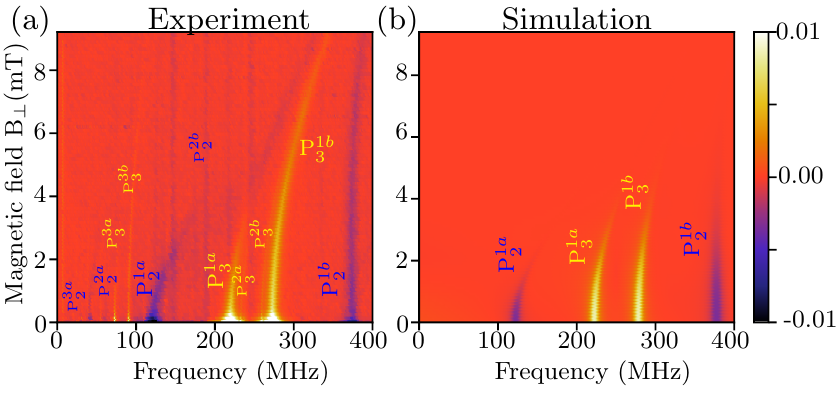}\caption{(a) Experimental and (b) simulated ODMR in 9 mT $\parallel$ c axis
along with sweeping magnetic field $\bot$ $c$-axis.}

\label{bper}
\end{figure}

The simulated ODMR of magnetic field small with c-axis enables the
double quantum transitions $d_{i=2,3}^{a(b)}$ . It is interesting
to see the effect of $B$ perpendicular to the c-axis. Figure$\;$\ref{bper}
(a) shows the ODMR recorded in different magnetic fields $\bot$ to
the $c$-axis in addition to a 9 mT magnetic field parallel to the
$c$-axis at room temperature. With increasing magnetic field strength
perpendicular to the $c$-axis, the signal amplitude of the peak $P_{3}^{1a}$
(due to spin transition 3/2$\leftrightarrow$1/2 ) falls rapidly compared
to the peak $P_{3}^{1b}$ (due to the spin transition -3/2$\leftrightarrow$-1/2).
At around 4.7 mT, the signal amplitude of the peak $P_{3}^{1a}$ is
almost zero, but the signal for $P_{3}^{1b}$ remains. With further
increase of the perpendicular magnetic field strength $P_{1a}^{3}$
amplitude becomes negative. The intensity pattern of the multi-photon
peaks is also similar to peaks $P_{i=2,3}^{1a(b)}$ . Figure$\;$\ref{bper}
(b) shows the simulated ODMR recorded in different magnetic fields
$\bot$ to the $c$-axis along with 9 mT magnetic field parallel to
the $c$-axis. The RF strength $\Omega_{1}$ used for both vacancies
was 2 MHz. In the simulations, we did not consider the effect of the
magnetic field on the excited and shelving states of the silicon-vacancy,
which can change the optical pumping scheme. Instead of equally populating
ground state spin levels $\pm1/2$, the spin level -1/2 gets slightly
more populated; this could be the reason for the difference between
the simulated and experimental intensity variation of peaks $P_{3}^{1b}$
and $P_{2}^{1b}$ with perpendicular magnetic.

\subsection{RF field with a component parallel to the $c$-axis}

\label{subsec:RF-field-parallel}

\begin{figure}
\includegraphics{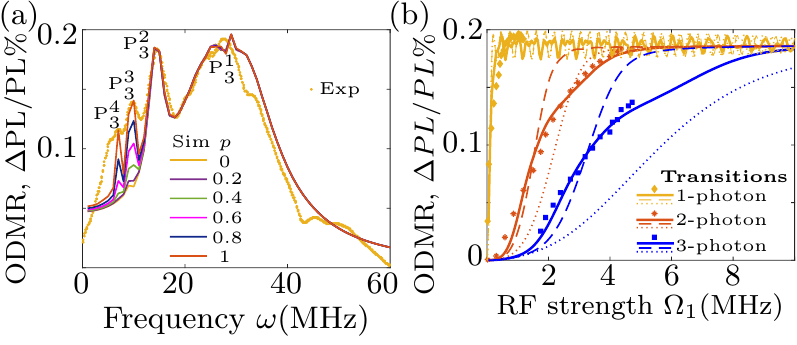}

\caption{(a) Experimental and simulated CW-ODMR plots for $V_{3}$ with the
RF Hamiltonian of Eq.$\;$\eqref{eq:hrfsz}. (b) Amplitudes of the
different peaks vs. RF strength. Diamonds, stars, and squares represent
experimentally measured data. Dotted, dashed, and solid curves are
the simulations obtained by solving Eq.$\;$\eqref{eq:lindblad} numerically
for different photons transitions with RF Hamiltonian given Eq.$\;$\eqref{eq:hrfsz}
with $p=0$, $=1$, and $=1$ along with consideration of the RF inhomogeneity,
respectively.}
\label{szsimexp-1}
\end{figure}

In the CW-ODMR experiments, we generate the RF field with a wire terminated
by a 50 W attenuator with 50 $\mathrm{\Omega}$ resistance. The resulting
RF Hamiltonian includes a component of the magnetic field parallel
to the $z$-axis ($c$-axis), which is usually ignored due to its
negligible effect on 1 RF photon absorption transitions. However,
with the relatively strong fields considered here, its effect can
not be neglected, as we show here. Figure$\;$\ref{szsimexp-1} (a)
shows the experimental and simulated CW-ODMR spectra for $V_{3}$.
The method used to simulate these plots is the same as explained in
the above Sec.$\;\ref{subsec:CW-ODMR}$ except the Hamiltonian used
is 
\begin{equation}
{\cal H}_{RF}(t)=\text{\ensuremath{\Omega_{1}}}cos\,(2\pi\omega t)\:(S_{x}+pS_{z}),\label{eq:hrfsz}
\end{equation}
where $\Omega_{1}$ =4.7 MHz and 0$\le$$p$$\le$1. The simulated
spectra were also renormalized to match the experimental signal. The
ODMR spectrum simulated with $p=1$ matches well with experimentally
measured spectrum. Figure$\;$\ref{szsimexp-1} (b) compares experimental
and simulated amplitudes of the different peaks vs. RF field strength.
Dotted, dashed and solid curves are the simulations obtained by solving
Eq.$\;$\eqref{eq:lindblad} numerically for different photons transitions
with RF Hamiltonian given Eq.$\;$\eqref{eq:hrfsz} with $p=0$, $=1$,
and $=1$ along with consideration of the RF inhomogeneity, respectively.
To account for the experimental RF inhomogeneity, signals with 0.5,
1 and 1.5 times $\Omega_{1}$, were calculated and added with weights
0.33, 0.34 and 0.33, respectively. We plot the experimental and simulated
data in the same plot using $\ensuremath{\Omega_{1}=\kappa\sqrt{RF\;POWER}}$
with $\kappa=0.25$ for 1-photon and $\approx1$ for 2- and 3-photon
transitions. Figure$\;$\ref{szsimexp-1} (b) shows that the $z$-axis
RF field strongly affects 2- and 3- RF photons absorption signals,
but its effect on 1- RF photons absorption peak is negligible. The
experimental data for amplitudes of the different peaks vs. RF field
strength matches with the solid simulated curves, which also account
for the inhomogeneity of the RF field.

\section{Discussion and Conclusion}

\label{sec:conc}

$V_{Si}^{-}$ in SiC is optically addressable and stores information
that can be manipulated using RF coherent control pulses. We studied
the silicon vacancies by applying the high-power RF pulses. In the
CW-ODMR experiments, we fitted the ODMR signal of 1, 2 and 3 RF photons
with the square root of the RF power. We applied the maximum possible
RF power with the RF amplifier and saturated the 1, 2 and 3 RF photons
peaks of the $V_{3}$ type of vacancy. But for the $V_{2}$ kind of
vacancy, we can only saturate the 1 RF photon absorption peak. The
1, 2 and 3 RF photons absorption peaks fitted well in Eq. 3 with the
$c$ value is =1, 2 and 3. The simulated ODMR Spectra and ODMR signal
vs. RF coupling strengths show similar dependence.

We measured the Rabi oscillations of $V_{3}$ type for the 1, 2, and
3 RF photon transitions, which fit well with RF coupling strength
$\Omega$= 9.1 MHZ for all oscillations, but the dephasing rates differed.
The 1 RF photon absorption transition dephase faster than 2 RF photons
absorption transition and 2 RF photons absorption transition dephase
faster than the 3 RF photons for the same RF power. It can be seen
from the CW-ODMR also that the linewidth of 2 RF photons absorption
transitions is smaller than the 1 RF photon absorption transitions.
We saw very sharp changes in the 3 RF photons absorption Rabi oscillation
simulations, but we did not see it in the experiment due to RF inhomogeneity.
Further, the simulations of the population dynamics with the $\sigma^{-}$
polarized RF field show that the three spin levels -3/2, 1/2, and
-1/2 were involved in the 2 and 3 RF photons absorption transitions.
The populations of spin levels -3/2 and 1/2 were exchanging in 2 RF
photons transitions, and for the 3 RF photons transitions, population
exchange was between the spin levels -3/2, 1/2, and -1/2. With linearly
polarized photons, the entire population flip happens between the
spin levels only with 1-photon absorption and for the 2- and 3-photon
it is RF strength depended. We successfully measured the FID of the
$V_{3}$ type by exciting it with 1, 2, and 3 RF photons.

The magnetic field lifts the degeneracy in the vacancy spin levels,
and we can see ODMR peaks for the different transitions. The ODMR
recorded in the magnetic field parallel to the $c$-axis clearly shows
these different RF photons transitions. The 1 RF photon absorption
is transition takes place between the spin levels $\pm$3/2$\leftrightarrow$$\pm$1/2
(with high ODMR intensity) and $\mp$3/2$\leftrightarrow$$\pm$1/2
(with low ODMR intensity). The 2 RF photons transitions are taking
place between the $\mp$3/2$\leftrightarrow$$\pm$1/2 and with very
low ODMR intensity in spin levels $\pm$3/2$\leftrightarrow$$\pm$1/2
also. In the transition $\pm$3/2$\leftrightarrow$$\pm$1/2 angular
momentum change is $\pm$2, so angular momentum is conserved for this
transition with two RF photons of the same polarization, wherein the
other is $\pm$1, and it is not conserved, giving low intensity. The
3 RF photons absorption transition is taking place between the $\pm$3/2$\leftrightarrow$$\pm$1/2.
The intensity of these 3 RF photons absorption peaks ($P_{3}^{3a}$
and $P_{3}^{3b}$) is higher than that of the 2 RF photons absorption
peaks ($P_{3}^{2a1}$ and $P_{3}^{2b1}$), i.e., transitions between
the spin levels $\pm$3/2$\leftrightarrow$$\pm$1/2. The higher intensity
in 3 RF photons transitions is due to angular momentum conservation,
and this could be possible if two out of three RF photons have the
same polarization. The ODMR recorded in 9 mT parallel to the c-axis
and along with the magnetic field perpendicular to the c-axis shows
the perpendicular magnetic field mixing more the transition -3/2$\leftrightarrow$1/2
as compared to the 3/2$\leftrightarrow$1/2. This effect could be
use to prepare a pseudo pure state of $V_{Si}^{-}$ spin ensemble.
Further, the multi-photon transitions depend not only on the amplitude
but also on the orientation of the RF field with respect to the symmetry
axis of the center.

So, in conclusion, 1, 2 and 3 photons transitions in the negatively
charged silicon-vacancy have been observed and characterized. Our
results improve the understanding of multi-photon absorption transitions
in silicon vacancies and help develop fast quantum coherent control
and other applications.

\section*{APPENDIX A: SAMPLE}

SiC crystals with a low content of background impurities were grown
from a synthesized source. Polycrystalline sources were synthesized
from semiconductor silicon and spectrally pure carbon. Polycrystalline
silicon and spectrally pure graphite in the form of powder were chosen
as the powder. Before the synthesis, the crucible and internal furnace
reinforcement were degassed at a temperature of 2200° C and a vacuum
of 10$^{-3}$ torrs for 2 hours by resistive heating growth machine,
after which the crucible was loaded with a mixture of carbon and silicon
powders in a stoichiometric ratio. The process of manufacturing silicon
carbide powder was carried out in a vacuum. To make the crucible,
graphite was used with a minimum content of background impurities,
such as Mersen 6516PT. In addition, it should be noted that all parts
of the crucible must be made of the same brand of graphite in order
to exclude the destruction of parts due to the difference in the coefficients
of thermal expansion of materials when the crucible was heated to
the synthesis temperature of the source (above 1600° C). As a seed,
a silicon carbide crystal of polytype 6H was used orientation on-axis.
Crystal growth parameters: temperature 2050° \textcyr{\CYRS}, argon
pressure about 2-5 torr, high-purity argon 99.9999\%, growth rate
about 150 $\mu$m/h. The micropore density in the grown crystal does
not exceed 3 cm$^{-2}$. For creating silicon vacancies, the crystal
was irradiated with electrons with a dose of $10^{18}$cm$^{-2}$
and an energy of 2 MeV at room temperature.

\section*{APPENDIX B: ODMR SETUP}

\begin{figure}
\includegraphics{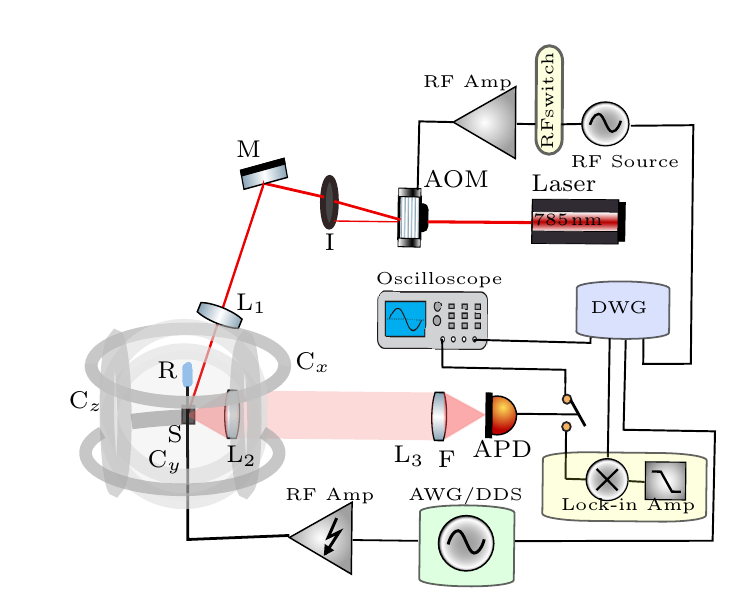}\caption{Experimental setup used for the ODMR experiments. An acoustic-optical
modulator (labeled AOM) is used to generate laser light pulses. Mirror,
lenses, iris and 850 nm long-pass filter are labeled M, L, I, and
F. An avalanche photodiode (APD) converts the PL signal into an electrical
signal. Three orthogonal sets of Helmholtz coils labeled C\protect\textsubscript{x},
C\protect\textsubscript{y}, and C\protect\textsubscript{z} are used
to apply a magnetic field in an arbitrary direction.}

\label{setup}
\end{figure}

Figure~\ref{setup} shows the setup used for the cw- and time-resolved
ODMR measurements, which is also used in our previous work$\;$\cite{singh-prb-20,singh-prb-21}.
We used a 785 nm diode laser as our light source (A laser diode (Thorlabs
LD785-SE400), a laser diode controller (LDC202C series) and a temperature
controller (TED 200C)). An acousto-optical modulator (NEC model OD8813A)
was used for creating the laser light pulses. For applying the static
magnetic field to the sample, we used three orthogonal coil-pairs.
The PL signal was recorded with an avalanche photodiode (APD) module
(C12703 series from Hamamatsu). The signal from the APD was recorded
with the USB oscilloscope card (PicoScope 2000 series) during pulse
mode ODMR experiments. For cw-ODMR, the signal from APD was recorded
with the lock-in (SRS model SR830 DSP). An RF source, we used a direct
digital synthesizer (DDS) AD9915 from Analog Devices. An RF signal
from the source was amplified using an RF amplifier (LZY-22+ from
mini circuits and ZHL-5W-1+). This amplified RF power feed to the
sample with a 50 $\mu$m diameter wire terminated with a 50-ohm resistor
via a 50 W attenuator. For pulsed ODMR experiments, we used an arbitrary
wave generator (AWG)(DAx14000 from Hunter Micro). An RF signal from
the source was amplified using an RF amplifier (LZY-22+ from mini
circuits). This amplified RF power was applied to the sample with
a Helmholtz-pair of RF coils with a diameter of 2.5 mm terminated
with a 50-ohm resistor. We used a digital word generator (DWG; SpinCore
PulseBlaster ESR-PRO PCI card) to generate TTL (transistor transistor
logic) pulses that trigger the laser RF pulses.

\section*{APPENDIX C: Lindblad Operators}

Lindblad operators used in the Lindblad equation Eq. \ref{eq:lindblad}
for including the optical pumping process of the $V_{3}$ type of
vacancy:

$L_{1}=\sqrt{\delta}\left(\begin{array}{cccc}
0 & 0 & 0 & 0\\
1 & 0 & 0 & 0\\
0 & 0 & 0 & 0\\
0 & 0 & 0 & 0
\end{array}\right)$, $L_{2}=\sqrt{\delta}\left(\begin{array}{cccc}
0 & 0 & 0 & 0\\
0 & 0 & 0 & 1\\
0 & 0 & 0 & 0\\
0 & 0 & 0 & 0
\end{array}\right)$, 

$L_{3}=\sqrt{\delta}\left(\begin{array}{cccc}
0 & 0 & 0 & 0\\
0 & 0 & 0 & 0\\
1 & 0 & 0 & 0\\
0 & 0 & 0 & 0
\end{array}\right)$, $L_{4}=\sqrt{\delta}\left(\begin{array}{cccc}
0 & 0 & 0 & 0\\
0 & 0 & 0 & 0\\
0 & 0 & 0 & 1\\
0 & 0 & 0 & 0
\end{array}\right)$, 

$L_{5}=\sqrt{\delta}\left(\begin{array}{cccc}
0 & 0 & 0 & 0\\
0 & 0 & 1 & 0\\
0 & 1 & 0 & 0\\
0 & 0 & 0 & 0
\end{array}\right)$,

Lindblad operators used for including the optical pumping process
of the $V_{2}$ type of vacancy:

$L_{1}=\sqrt{\delta}\left(\begin{array}{cccc}
0 & 1 & 0 & 0\\
0 & 0 & 0 & 0\\
0 & 0 & 0 & 0\\
0 & 0 & 0 & 0
\end{array}\right)$, $L_{2}=\sqrt{\delta}\left(\begin{array}{cccc}
0 & 0 & 0 & 0\\
0 & 0 & 0 & 0\\
0 & 0 & 0 & 0\\
0 & 1 & 0 & 0
\end{array}\right)$,

$L_{3}=\sqrt{\delta}\left(\begin{array}{cccc}
0 & 0 & 1 & 0\\
0 & 0 & 0 & 0\\
0 & 0 & 0 & 0\\
0 & 0 & 0 & 0
\end{array}\right)$, $L_{4}=\sqrt{\delta}\left(\begin{array}{cccc}
0 & 0 & 0 & 0\\
0 & 0 & 0 & 0\\
0 & 0 & 0 & 0\\
0 & 0 & 1 & 0
\end{array}\right)$,

$L_{5}=\sqrt{\delta}\left(\begin{array}{cccc}
0 & 0 & 0 & 1\\
0 & 0 & 0 & 0\\
0 & 0 & 0 & 0\\
1 & 0 & 0 & 0
\end{array}\right)$.

Lindblad operators for including the relaxation contribution $L_{6}=\sqrt{2\beta}\:S_{z}$,
$L_{7}=\sqrt{2\alpha}\:S_{x}$, where $S_{z}=\frac{1}{2}\left(\begin{array}{cccc}
3 & 0 & 0 & 0\\
0 & 1 & 0 & 0\\
0 & 0 & -1 & 0\\
0 & 0 & 0 & -3
\end{array}\right)$ and $S_{x}=\frac{1}{2}\left(\begin{array}{cccc}
0 & \sqrt{3} & 0 & 0\\
\sqrt{3} & 0 & 2 & 0\\
0 & 2 & 0 & \sqrt{3}\\
0 & 0 & \sqrt{3} & 0
\end{array}\right)$.

The values of $k_{i}$ are used to solve Eq.\ref{eq:lindblad} with
the Runge-Kutta 4 method.
\begin{itemize}
\item $\ensuremath{k_{1}=-i({\cal H}_{t}(t_{n}).\rho(n)-\rho(n).{\cal H}_{t}(t_{n}))+\sum_{\alpha,\beta,\delta}L\ensuremath{_{i}}^{\dagger}.\rho(n).L\ensuremath{_{i}}-\frac{1}{2}(L_{i}^{\dagger}L_{i}.\rho(n)+\rho(n).L_{i}^{\dagger}L_{i});}$
\item $k_{2}=-i({\cal H}_{t}(\frac{\ensuremath{\Delta}t}{2}+t_{n}).(\frac{\ensuremath{\Delta}tk_{1}}{2}+\rho_{n})-(\frac{\text{\ensuremath{\Delta}}tk_{1}}{2}+\rho_{n}).{\cal H}_{t}(\frac{\text{\ensuremath{\Delta}}t}{2}+t_{n}))+\sum_{\alpha,\beta,\delta}L\ensuremath{_{i}}.(\frac{\Delta tk_{1}}{2}+\rho_{n}).L\ensuremath{_{i}}^{\dagger}-\frac{1}{2}((\frac{\Delta tk_{1}}{2}+\rho_{n}).\text{\ensuremath{L_{i}^{\dagger}L_{i}}}+\text{\ensuremath{L_{i}^{\dagger}L_{i}}}.(\frac{\Delta tk_{1}}{2}+\rho_{n}));$
\item $k_{3}=-i({\cal H}_{t}(\frac{\ensuremath{\Delta}t}{2}+t_{n}).(\frac{\Delta tk_{2}}{2}+\rho_{n})-(\frac{\text{\ensuremath{\Delta}}tk_{2}}{2}+\rho_{n}).{\cal H}_{t}(\frac{\text{\ensuremath{\Delta}}t}{2}+t_{n}))+\sum_{\alpha,\beta,\delta}L\ensuremath{_{i}}.(\frac{\Delta tk_{2}}{2}+\rho_{n}).L\ensuremath{_{i}}^{\dagger}-\frac{1}{2}((\frac{\Delta tk_{2}}{2}+\rho_{n}).\text{\ensuremath{L_{i}^{\dagger}L_{i}}}+\text{\ensuremath{L_{i}^{\dagger}L_{i}}}.(\frac{\Delta tk_{2}}{2}+\rho_{n}));$
\item $\ensuremath{k_{4}=-i({\cal H}_{t}(\ensuremath{\Delta t}+t_{n}).(\ensuremath{\Delta}t\:k_{3}+t_{n})-(\ensuremath{\Delta}t\:k_{3}+\rho_{n}).{\cal H}_{t}(\ensuremath{\Delta}t+t_{n}))+\sum_{\alpha,\beta,\delta}L\ensuremath{_{i}^{\dagger}}.(\ensuremath{\Delta}t\,k_{3}+\rho_{n}).L\ensuremath{_{i}}-\frac{1}{2}((\ensuremath{\Delta\,}tk_{3}+\rho_{n}).\text{\ensuremath{L_{i}^{\dagger}L_{i}}}+L_{i}^{\dagger}L_{i}.(\ensuremath{\Delta}t\:k_{3}+\rho_{n}));}$
\end{itemize}
$\ensuremath{\rho_{n+1}=\rho_{n}+\frac{1}{6}\ensuremath{\Delta}t(k_{1}+2k_{2}+2k_{3}+k_{4})}$;

\begin{figure}
\includegraphics[scale=1.5]{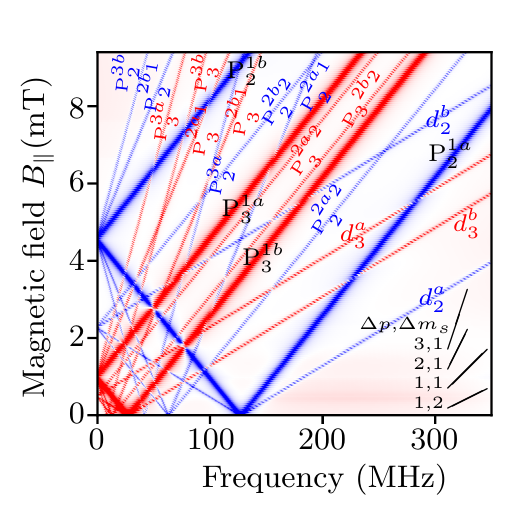}

\caption{Plot showing different possible transitions for the $V_{1}/V_{3}$
and $V_{2}$ for a range of magnetic fields B $\parallel$ c axis,
calculated using eigenvalues of Hamiltonian given in Eq.$\;$\eqref{eq:hamiltonian}.
Peaks labeled with $P_{2}^{1a(b)}$ ($P_{3}^{1a(b)}$), $P_{2}^{2a(b)}$
($P_{3}^{2a(b)}$ ) and $P_{2}^{3a(b)}$ ($P_{3}^{3a(b)}$ ) are 1,
2 and 3, RF photons absorption peaks due to the single quantum spin
transition $\Delta$$m_{s}=1$, where peaks $d_{2}^{a(b)}$($d_{3}^{a(b)}$)
are 1 RF photon absorption peaks due to double quantum spin transition
$\Delta$$m_{s}=2$ for $V_{2}$ ( $V_{3}$ ) type of $V_{Si}^{-}$.
Slope of different transitions are depends on the number RF photon
absorb $\Delta p$ and the change in the spin quantum number $\Delta m_{s}$
during the transitions.}

\label{trasitionfromeig}
\end{figure}

Figure$\;$\ref{trasitionfromeig} shows different possible transitions
for the $V_{1}/V_{3}$ and $V_{2}$ for a range of magnetic fields
B $\parallel$ c axis, calculated using eigenvalues of Hamiltonian
given in Eq.$\;$\eqref{eq:hamiltonian}. Red and blue color lines
represent the possible 1-, 2-, 3- RF photon absorption transitions.

\begin{acknowledgments}
This work was supported by the Deutsche Forschungsgemeinschaft in
the frame of the ICRC TRR 160 (Project No. C7) and by RFBR, project
number 19-52-12058.
\end{acknowledgments}


\begin{thebibliography}{30}%
\makeatletter
\providecommand \@ifxundefined [1]{%
 \@ifx{#1\undefined}
}%
\providecommand \@ifnum [1]{%
 \ifnum #1\expandafter \@firstoftwo
 \else \expandafter \@secondoftwo
 \fi
}%
\providecommand \@ifx [1]{%
 \ifx #1\expandafter \@firstoftwo
 \else \expandafter \@secondoftwo
 \fi
}%
\providecommand \natexlab [1]{#1}%
\providecommand \enquote  [1]{``#1''}%
\providecommand \bibnamefont  [1]{#1}%
\providecommand \bibfnamefont [1]{#1}%
\providecommand \citenamefont [1]{#1}%
\providecommand \href@noop [0]{\@secondoftwo}%
\providecommand \href [0]{\begingroup \@sanitize@url \@href}%
\providecommand \@href[1]{\@@startlink{#1}\@@href}%
\providecommand \@@href[1]{\endgroup#1\@@endlink}%
\providecommand \@sanitize@url [0]{\catcode `\\12\catcode `\$12\catcode
  `\&12\catcode `\#12\catcode `\^12\catcode `\_12\catcode `\%12\relax}%
\providecommand \@@startlink[1]{}%
\providecommand \@@endlink[0]{}%
\providecommand \url  [0]{\begingroup\@sanitize@url \@url }%
\providecommand \@url [1]{\endgroup\@href {#1}{\urlprefix }}%
\providecommand \urlprefix  [0]{URL }%
\providecommand \Eprint [0]{\href }%
\providecommand \doibase [0]{http://dx.doi.org/}%
\providecommand \selectlanguage [0]{\@gobble}%
\providecommand \bibinfo  [0]{\@secondoftwo}%
\providecommand \bibfield  [0]{\@secondoftwo}%
\providecommand \translation [1]{[#1]}%
\providecommand \BibitemOpen [0]{}%
\providecommand \bibitemStop [0]{}%
\providecommand \bibitemNoStop [0]{.\EOS\space}%
\providecommand \EOS [0]{\spacefactor3000\relax}%
\providecommand \BibitemShut  [1]{\csname bibitem#1\endcsname}%
\let\auto@bib@innerbib\@empty
\bibitem [{\citenamefont {Singh}\ \emph
  {et~al.}(2020{\natexlab{a}})\citenamefont {Singh}, \citenamefont {Anisimov},
  \citenamefont {Baranov},\ and\ \citenamefont {Suter}}]{singh2020optical}%
  \BibitemOpen
  \bibfield  {author} {\bibinfo {author} {\bibfnamefont {H.}~\bibnamefont
  {Singh}}, \bibinfo {author} {\bibfnamefont {A.~N.}\ \bibnamefont {Anisimov}},
  \bibinfo {author} {\bibfnamefont {P.~G.}\ \bibnamefont {Baranov}}, \ and\
  \bibinfo {author} {\bibfnamefont {D.}~\bibnamefont {Suter}},\ }\href@noop {}
  {\bibfield  {journal} {\bibinfo  {journal} {arXiv preprint arXiv:2007.08516}\
  } (\bibinfo {year} {2020}{\natexlab{a}})}\BibitemShut {NoStop}%
\bibitem [{\citenamefont {Soltamov}\ \emph {et~al.}(2019)\citenamefont
  {Soltamov}, \citenamefont {Kasper}, \citenamefont {Poshakinskiy},
  \citenamefont {Anisimov}, \citenamefont {Mokhov}, \citenamefont {Sperlich},
  \citenamefont {Tarasenko}, \citenamefont {Baranov}, \citenamefont
  {Astakhov},\ and\ \citenamefont {Dyakonov}}]{soltamov-naturecom-19}%
  \BibitemOpen
  \bibfield  {author} {\bibinfo {author} {\bibfnamefont {V.~A.}\ \bibnamefont
  {Soltamov}}, \bibinfo {author} {\bibfnamefont {C.}~\bibnamefont {Kasper}},
  \bibinfo {author} {\bibfnamefont {A.~V.}\ \bibnamefont {Poshakinskiy}},
  \bibinfo {author} {\bibfnamefont {A.~N.}\ \bibnamefont {Anisimov}}, \bibinfo
  {author} {\bibfnamefont {E.~N.}\ \bibnamefont {Mokhov}}, \bibinfo {author}
  {\bibfnamefont {A.}~\bibnamefont {Sperlich}}, \bibinfo {author}
  {\bibfnamefont {S.~A.}\ \bibnamefont {Tarasenko}}, \bibinfo {author}
  {\bibfnamefont {P.~G.}\ \bibnamefont {Baranov}}, \bibinfo {author}
  {\bibfnamefont {G.~V.}\ \bibnamefont {Astakhov}}, \ and\ \bibinfo {author}
  {\bibfnamefont {V.}~\bibnamefont {Dyakonov}},\ }\href {\doibase
  10.1038/s41467-019-09429-x} {\bibfield  {journal} {\bibinfo  {journal}
  {Nature Communications}\ }\textbf {\bibinfo {volume} {10}},\ \bibinfo {pages}
  {1678} (\bibinfo {year} {2019})}\BibitemShut {NoStop}%
\bibitem [{\citenamefont {Widmann}\ \emph {et~al.}(2015)\citenamefont
  {Widmann}, \citenamefont {Lee}, \citenamefont {Rendler}, \citenamefont {Son},
  \citenamefont {Fedder}, \citenamefont {Paik}, \citenamefont {Yang},
  \citenamefont {Zhao}, \citenamefont {Yang}, \citenamefont {Booker} \emph
  {et~al.}}]{widmann-nature-14}%
  \BibitemOpen
  \bibfield  {author} {\bibinfo {author} {\bibfnamefont {M.}~\bibnamefont
  {Widmann}}, \bibinfo {author} {\bibfnamefont {S.-Y.}\ \bibnamefont {Lee}},
  \bibinfo {author} {\bibfnamefont {T.}~\bibnamefont {Rendler}}, \bibinfo
  {author} {\bibfnamefont {N.~T.}\ \bibnamefont {Son}}, \bibinfo {author}
  {\bibfnamefont {H.}~\bibnamefont {Fedder}}, \bibinfo {author} {\bibfnamefont
  {S.}~\bibnamefont {Paik}}, \bibinfo {author} {\bibfnamefont {L.-P.}\
  \bibnamefont {Yang}}, \bibinfo {author} {\bibfnamefont {N.}~\bibnamefont
  {Zhao}}, \bibinfo {author} {\bibfnamefont {S.}~\bibnamefont {Yang}}, \bibinfo
  {author} {\bibfnamefont {I.}~\bibnamefont {Booker}},  \emph {et~al.},\ }\href
  {\doibase 10.1038/nmat4145} {\bibfield  {journal} {\bibinfo  {journal}
  {Nature materials}\ }\textbf {\bibinfo {volume} {14}},\ \bibinfo {pages}
  {164} (\bibinfo {year} {2015})}\BibitemShut {NoStop}%
\bibitem [{\citenamefont {Lukin}\ \emph {et~al.}(2020)\citenamefont {Lukin},
  \citenamefont {Guidry},\ and\ \citenamefont {Vu\ifmmode \check{c}\else
  \v{c}\fi{}kovi\ifmmode~\acute{c}\else \'{c}\fi{}}}]{lukin-prxq-20}%
  \BibitemOpen
  \bibfield  {author} {\bibinfo {author} {\bibfnamefont {D.~M.}\ \bibnamefont
  {Lukin}}, \bibinfo {author} {\bibfnamefont {M.~A.}\ \bibnamefont {Guidry}}, \
  and\ \bibinfo {author} {\bibfnamefont {J.}~\bibnamefont {Vu\ifmmode
  \check{c}\else \v{c}\fi{}kovi\ifmmode~\acute{c}\else \'{c}\fi{}}},\ }\href
  {\doibase 10.1103/PRXQuantum.1.020102} {\bibfield  {journal} {\bibinfo
  {journal} {PRX Quantum}\ }\textbf {\bibinfo {volume} {1}},\ \bibinfo {pages}
  {020102} (\bibinfo {year} {2020})}\BibitemShut {NoStop}%
\bibitem [{\citenamefont {Baranov}\ \emph {et~al.}(2011)\citenamefont
  {Baranov}, \citenamefont {Bundakova}, \citenamefont {Soltamova},
  \citenamefont {Orlinskii}, \citenamefont {Borovykh}, \citenamefont
  {Zondervan}, \citenamefont {Verberk},\ and\ \citenamefont
  {Schmidt}}]{baranov-prb-11}%
  \BibitemOpen
  \bibfield  {author} {\bibinfo {author} {\bibfnamefont {P.~G.}\ \bibnamefont
  {Baranov}}, \bibinfo {author} {\bibfnamefont {A.~P.}\ \bibnamefont
  {Bundakova}}, \bibinfo {author} {\bibfnamefont {A.~A.}\ \bibnamefont
  {Soltamova}}, \bibinfo {author} {\bibfnamefont {S.~B.}\ \bibnamefont
  {Orlinskii}}, \bibinfo {author} {\bibfnamefont {I.~V.}\ \bibnamefont
  {Borovykh}}, \bibinfo {author} {\bibfnamefont {R.}~\bibnamefont {Zondervan}},
  \bibinfo {author} {\bibfnamefont {R.}~\bibnamefont {Verberk}}, \ and\
  \bibinfo {author} {\bibfnamefont {J.}~\bibnamefont {Schmidt}},\ }\href
  {\doibase 10.1103/PhysRevB.83.125203} {\bibfield  {journal} {\bibinfo
  {journal} {Phys. Rev. B}\ }\textbf {\bibinfo {volume} {83}},\ \bibinfo
  {pages} {125203} (\bibinfo {year} {2011})}\BibitemShut {NoStop}%
\bibitem [{\citenamefont {Christle}\ \emph {et~al.}(2015)\citenamefont
  {Christle}, \citenamefont {Falk}, \citenamefont {Andrich}, \citenamefont
  {Klimov}, \citenamefont {Hassan}, \citenamefont {Son}, \citenamefont
  {Janz\'en}, \citenamefont {Ohshima},\ and\ \citenamefont
  {Awschalom}}]{christle-nature-14}%
  \BibitemOpen
  \bibfield  {author} {\bibinfo {author} {\bibfnamefont {D.~J.}\ \bibnamefont
  {Christle}}, \bibinfo {author} {\bibfnamefont {A.~L.}\ \bibnamefont {Falk}},
  \bibinfo {author} {\bibfnamefont {P.}~\bibnamefont {Andrich}}, \bibinfo
  {author} {\bibfnamefont {P.~V.}\ \bibnamefont {Klimov}}, \bibinfo {author}
  {\bibfnamefont {J.~U.}\ \bibnamefont {Hassan}}, \bibinfo {author}
  {\bibfnamefont {N.~T.}\ \bibnamefont {Son}}, \bibinfo {author} {\bibfnamefont
  {E.}~\bibnamefont {Janz\'en}}, \bibinfo {author} {\bibfnamefont
  {T.}~\bibnamefont {Ohshima}}, \ and\ \bibinfo {author} {\bibfnamefont
  {D.~D.}\ \bibnamefont {Awschalom}},\ }\href
  {https://www.nature.com/articles/nmat4144#supplementary-information}
  {\bibfield  {journal} {\bibinfo  {journal} {Nature materials}\ }\textbf
  {\bibinfo {volume} {14}},\ \bibinfo {pages} {160} (\bibinfo {year}
  {2015})}\BibitemShut {NoStop}%
\bibitem [{\citenamefont {Falk}\ \emph {et~al.}(2013)\citenamefont {Falk},
  \citenamefont {Buckley}, \citenamefont {Calusine}, \citenamefont {Koehl},
  \citenamefont {Dobrovitski}, \citenamefont {Politi}, \citenamefont {Zorman},
  \citenamefont {Feng},\ and\ \citenamefont {Awschalom}}]{falk-nature-13}%
  \BibitemOpen
  \bibfield  {author} {\bibinfo {author} {\bibfnamefont {A.~L.}\ \bibnamefont
  {Falk}}, \bibinfo {author} {\bibfnamefont {B.~B.}\ \bibnamefont {Buckley}},
  \bibinfo {author} {\bibfnamefont {G.}~\bibnamefont {Calusine}}, \bibinfo
  {author} {\bibfnamefont {W.~F.}\ \bibnamefont {Koehl}}, \bibinfo {author}
  {\bibfnamefont {V.~V.}\ \bibnamefont {Dobrovitski}}, \bibinfo {author}
  {\bibfnamefont {A.}~\bibnamefont {Politi}}, \bibinfo {author} {\bibfnamefont
  {C.~A.}\ \bibnamefont {Zorman}}, \bibinfo {author} {\bibfnamefont {P.~X.-L.}\
  \bibnamefont {Feng}}, \ and\ \bibinfo {author} {\bibfnamefont {D.~D.}\
  \bibnamefont {Awschalom}},\ }\href {\doibase 10.1038/ncomms2854} {\bibfield
  {journal} {\bibinfo  {journal} {Nature communications}\ }\textbf {\bibinfo
  {volume} {4}},\ \bibinfo {pages} {1} (\bibinfo {year} {2013})}\BibitemShut
  {NoStop}%
\bibitem [{\citenamefont {Kraus}\ \emph {et~al.}(2014)\citenamefont {Kraus},
  \citenamefont {Soltamov}, \citenamefont {Riedel}, \citenamefont {V{\"a}th},
  \citenamefont {Fuchs}, \citenamefont {Sperlich}, \citenamefont {Baranov},
  \citenamefont {Dyakonov},\ and\ \citenamefont {Astakhov}}]{kraus-nature-13}%
  \BibitemOpen
  \bibfield  {author} {\bibinfo {author} {\bibfnamefont {H.}~\bibnamefont
  {Kraus}}, \bibinfo {author} {\bibfnamefont {V.}~\bibnamefont {Soltamov}},
  \bibinfo {author} {\bibfnamefont {D.}~\bibnamefont {Riedel}}, \bibinfo
  {author} {\bibfnamefont {S.}~\bibnamefont {V{\"a}th}}, \bibinfo {author}
  {\bibfnamefont {F.}~\bibnamefont {Fuchs}}, \bibinfo {author} {\bibfnamefont
  {A.}~\bibnamefont {Sperlich}}, \bibinfo {author} {\bibfnamefont
  {P.}~\bibnamefont {Baranov}}, \bibinfo {author} {\bibfnamefont
  {V.}~\bibnamefont {Dyakonov}}, \ and\ \bibinfo {author} {\bibfnamefont
  {G.}~\bibnamefont {Astakhov}},\ }\href@noop {} {\bibfield  {journal}
  {\bibinfo  {journal} {Nature Physics}\ }\textbf {\bibinfo {volume} {10}},\
  \bibinfo {pages} {157} (\bibinfo {year} {2014})}\BibitemShut {NoStop}%
\bibitem [{\citenamefont {Singh}\ \emph {et~al.}(2021)\citenamefont {Singh},
  \citenamefont {Anisimov}, \citenamefont {Breev}, \citenamefont {Baranov},\
  and\ \citenamefont {Suter}}]{singh-prb-21}%
  \BibitemOpen
  \bibfield  {author} {\bibinfo {author} {\bibfnamefont {H.}~\bibnamefont
  {Singh}}, \bibinfo {author} {\bibfnamefont {A.~N.}\ \bibnamefont {Anisimov}},
  \bibinfo {author} {\bibfnamefont {I.~D.}\ \bibnamefont {Breev}}, \bibinfo
  {author} {\bibfnamefont {P.~G.}\ \bibnamefont {Baranov}}, \ and\ \bibinfo
  {author} {\bibfnamefont {D.}~\bibnamefont {Suter}},\ }\href {\doibase
  10.1103/PhysRevB.103.104103} {\bibfield  {journal} {\bibinfo  {journal}
  {Phys. Rev. B}\ }\textbf {\bibinfo {volume} {103}},\ \bibinfo {pages}
  {104103} (\bibinfo {year} {2021})}\BibitemShut {NoStop}%
\bibitem [{\citenamefont {Singh}\ \emph
  {et~al.}(2020{\natexlab{b}})\citenamefont {Singh}, \citenamefont {Anisimov},
  \citenamefont {Nagalyuk}, \citenamefont {Mokhov}, \citenamefont {Baranov},\
  and\ \citenamefont {Suter}}]{singh-prb-20}%
  \BibitemOpen
  \bibfield  {author} {\bibinfo {author} {\bibfnamefont {H.}~\bibnamefont
  {Singh}}, \bibinfo {author} {\bibfnamefont {A.~N.}\ \bibnamefont {Anisimov}},
  \bibinfo {author} {\bibfnamefont {S.~S.}\ \bibnamefont {Nagalyuk}}, \bibinfo
  {author} {\bibfnamefont {E.~N.}\ \bibnamefont {Mokhov}}, \bibinfo {author}
  {\bibfnamefont {P.~G.}\ \bibnamefont {Baranov}}, \ and\ \bibinfo {author}
  {\bibfnamefont {D.}~\bibnamefont {Suter}},\ }\href {\doibase
  10.1103/PhysRevB.101.134110} {\bibfield  {journal} {\bibinfo  {journal}
  {Phys. Rev. B}\ }\textbf {\bibinfo {volume} {101}},\ \bibinfo {pages}
  {134110} (\bibinfo {year} {2020}{\natexlab{b}})}\BibitemShut {NoStop}%
\bibitem [{\citenamefont {Soltamov}\ \emph {et~al.}(2021)\citenamefont
  {Soltamov}, \citenamefont {Yavkin}, \citenamefont {Anisimov}, \citenamefont
  {Singh}, \citenamefont {Bundakova}, \citenamefont {Mamin}, \citenamefont
  {Orlinskii}, \citenamefont {Mokhov}, \citenamefont {Suter},\ and\
  \citenamefont {Baranov}}]{soltamov-prb-21}%
  \BibitemOpen
  \bibfield  {author} {\bibinfo {author} {\bibfnamefont {V.~A.}\ \bibnamefont
  {Soltamov}}, \bibinfo {author} {\bibfnamefont {B.~V.}\ \bibnamefont
  {Yavkin}}, \bibinfo {author} {\bibfnamefont {A.~N.}\ \bibnamefont
  {Anisimov}}, \bibinfo {author} {\bibfnamefont {H.}~\bibnamefont {Singh}},
  \bibinfo {author} {\bibfnamefont {A.~P.}\ \bibnamefont {Bundakova}}, \bibinfo
  {author} {\bibfnamefont {G.~V.}\ \bibnamefont {Mamin}}, \bibinfo {author}
  {\bibfnamefont {S.~B.}\ \bibnamefont {Orlinskii}}, \bibinfo {author}
  {\bibfnamefont {E.~N.}\ \bibnamefont {Mokhov}}, \bibinfo {author}
  {\bibfnamefont {D.}~\bibnamefont {Suter}}, \ and\ \bibinfo {author}
  {\bibfnamefont {P.~G.}\ \bibnamefont {Baranov}},\ }\href {\doibase
  10.1103/PhysRevB.103.195201} {\bibfield  {journal} {\bibinfo  {journal}
  {Phys. Rev. B}\ }\textbf {\bibinfo {volume} {103}},\ \bibinfo {pages}
  {195201} (\bibinfo {year} {2021})}\BibitemShut {NoStop}%
\bibitem [{\citenamefont {S\"orman}\ \emph {et~al.}(2000)\citenamefont
  {S\"orman}, \citenamefont {Son}, \citenamefont {Chen}, \citenamefont
  {Kordina}, \citenamefont {Hallin},\ and\ \citenamefont
  {Janz\'en}}]{sorman-prb-00}%
  \BibitemOpen
  \bibfield  {author} {\bibinfo {author} {\bibfnamefont {E.}~\bibnamefont
  {S\"orman}}, \bibinfo {author} {\bibfnamefont {N.~T.}\ \bibnamefont {Son}},
  \bibinfo {author} {\bibfnamefont {W.~M.}\ \bibnamefont {Chen}}, \bibinfo
  {author} {\bibfnamefont {O.}~\bibnamefont {Kordina}}, \bibinfo {author}
  {\bibfnamefont {C.}~\bibnamefont {Hallin}}, \ and\ \bibinfo {author}
  {\bibfnamefont {E.}~\bibnamefont {Janz\'en}},\ }\href {\doibase
  10.1103/PhysRevB.61.2613} {\bibfield  {journal} {\bibinfo  {journal} {Phys.
  Rev. B}\ }\textbf {\bibinfo {volume} {61}},\ \bibinfo {pages} {2613}
  (\bibinfo {year} {2000})}\BibitemShut {NoStop}%
\bibitem [{\citenamefont {Biktagirov}\ \emph {et~al.}(2018)\citenamefont
  {Biktagirov}, \citenamefont {Schmidt}, \citenamefont {Gerstmann},
  \citenamefont {Yavkin}, \citenamefont {Orlinskii}, \citenamefont {Baranov},
  \citenamefont {Dyakonov},\ and\ \citenamefont
  {Soltamov}}]{biktagirov-prb-18}%
  \BibitemOpen
  \bibfield  {author} {\bibinfo {author} {\bibfnamefont {T.}~\bibnamefont
  {Biktagirov}}, \bibinfo {author} {\bibfnamefont {W.~G.}\ \bibnamefont
  {Schmidt}}, \bibinfo {author} {\bibfnamefont {U.}~\bibnamefont {Gerstmann}},
  \bibinfo {author} {\bibfnamefont {B.}~\bibnamefont {Yavkin}}, \bibinfo
  {author} {\bibfnamefont {S.}~\bibnamefont {Orlinskii}}, \bibinfo {author}
  {\bibfnamefont {P.}~\bibnamefont {Baranov}}, \bibinfo {author} {\bibfnamefont
  {V.}~\bibnamefont {Dyakonov}}, \ and\ \bibinfo {author} {\bibfnamefont
  {V.}~\bibnamefont {Soltamov}},\ }\href {\doibase 10.1103/PhysRevB.98.195204}
  {\bibfield  {journal} {\bibinfo  {journal} {Phys. Rev. B}\ }\textbf {\bibinfo
  {volume} {98}},\ \bibinfo {pages} {195204} (\bibinfo {year}
  {2018})}\BibitemShut {NoStop}%
\bibitem [{\citenamefont {Davidsson}\ \emph {et~al.}(2019)\citenamefont
  {Davidsson}, \citenamefont {Ivády}, \citenamefont {Armiento}, \citenamefont
  {Ohshima}, \citenamefont {Son}, \citenamefont {Gali},\ and\ \citenamefont
  {Abrikosov}}]{davidsson-apl-19}%
  \BibitemOpen
  \bibfield  {author} {\bibinfo {author} {\bibfnamefont {J.}~\bibnamefont
  {Davidsson}}, \bibinfo {author} {\bibfnamefont {V.}~\bibnamefont {Ivády}},
  \bibinfo {author} {\bibfnamefont {R.}~\bibnamefont {Armiento}}, \bibinfo
  {author} {\bibfnamefont {T.}~\bibnamefont {Ohshima}}, \bibinfo {author}
  {\bibfnamefont {N.~T.}\ \bibnamefont {Son}}, \bibinfo {author} {\bibfnamefont
  {A.}~\bibnamefont {Gali}}, \ and\ \bibinfo {author} {\bibfnamefont {I.~A.}\
  \bibnamefont {Abrikosov}},\ }\href {\doibase 10.1063/1.5083031} {\bibfield
  {journal} {\bibinfo  {journal} {Applied Physics Letters}\ }\textbf {\bibinfo
  {volume} {114}},\ \bibinfo {pages} {112107} (\bibinfo {year} {2019})},\
  \Eprint {http://arxiv.org/abs/https://doi.org/10.1063/1.5083031}
  {https://doi.org/10.1063/1.5083031} \BibitemShut {NoStop}%
\bibitem [{\citenamefont {Breev}\ \emph {et~al.}(2021)\citenamefont {Breev},
  \citenamefont {Shang}, \citenamefont {Poshakinskiy}, \citenamefont {Singh},
  \citenamefont {Berencen}, \citenamefont {Hollenbach}, \citenamefont
  {Nagalyuk}, \citenamefont {Mokhov}, \citenamefont {Babunts}, \citenamefont
  {Baranov}, \citenamefont {Suter}, \citenamefont {Tarasenko}, \citenamefont
  {Astakhov},\ and\ \citenamefont {Anisimov}}]{breev2021inverted}%
  \BibitemOpen
  \bibfield  {author} {\bibinfo {author} {\bibfnamefont {I.~D.}\ \bibnamefont
  {Breev}}, \bibinfo {author} {\bibfnamefont {Z.}~\bibnamefont {Shang}},
  \bibinfo {author} {\bibfnamefont {A.~V.}\ \bibnamefont {Poshakinskiy}},
  \bibinfo {author} {\bibfnamefont {H.}~\bibnamefont {Singh}}, \bibinfo
  {author} {\bibfnamefont {Y.}~\bibnamefont {Berencen}}, \bibinfo {author}
  {\bibfnamefont {M.}~\bibnamefont {Hollenbach}}, \bibinfo {author}
  {\bibfnamefont {S.~S.}\ \bibnamefont {Nagalyuk}}, \bibinfo {author}
  {\bibfnamefont {E.~N.}\ \bibnamefont {Mokhov}}, \bibinfo {author}
  {\bibfnamefont {R.~A.}\ \bibnamefont {Babunts}}, \bibinfo {author}
  {\bibfnamefont {P.~G.}\ \bibnamefont {Baranov}}, \bibinfo {author}
  {\bibfnamefont {D.}~\bibnamefont {Suter}}, \bibinfo {author} {\bibfnamefont
  {S.~A.}\ \bibnamefont {Tarasenko}}, \bibinfo {author} {\bibfnamefont {G.~V.}\
  \bibnamefont {Astakhov}}, \ and\ \bibinfo {author} {\bibfnamefont {A.~N.}\
  \bibnamefont {Anisimov}},\ }\href@noop {} {\enquote {\bibinfo {title}
  {Inverted fine structure of a 6h-sic qubit enabling robust spin-photon
  interface},}\ } (\bibinfo {year} {2021}),\ \Eprint
  {http://arxiv.org/abs/2107.06989} {arXiv:2107.06989 [quant-ph]} \BibitemShut
  {NoStop}%
\bibitem [{\citenamefont {Nathan}\ \emph {et~al.}(1985)\citenamefont {Nathan},
  \citenamefont {Guenther},\ and\ \citenamefont {Mitra}}]{nathan1985review}%
  \BibitemOpen
  \bibfield  {author} {\bibinfo {author} {\bibfnamefont {V.}~\bibnamefont
  {Nathan}}, \bibinfo {author} {\bibfnamefont {A.~H.}\ \bibnamefont
  {Guenther}}, \ and\ \bibinfo {author} {\bibfnamefont {S.~S.}\ \bibnamefont
  {Mitra}},\ }\href@noop {} {\bibfield  {journal} {\bibinfo  {journal} {JOSA
  B}\ }\textbf {\bibinfo {volume} {2}},\ \bibinfo {pages} {294} (\bibinfo
  {year} {1985})}\BibitemShut {NoStop}%
\bibitem [{\citenamefont {Wallis}\ \emph {et~al.}(1974)\citenamefont {Wallis},
  \citenamefont {Brion}, \citenamefont {Burstein},\ and\ \citenamefont
  {Hartstein}}]{wallis-prb-74}%
  \BibitemOpen
  \bibfield  {author} {\bibinfo {author} {\bibfnamefont {R.~F.}\ \bibnamefont
  {Wallis}}, \bibinfo {author} {\bibfnamefont {J.~J.}\ \bibnamefont {Brion}},
  \bibinfo {author} {\bibfnamefont {E.}~\bibnamefont {Burstein}}, \ and\
  \bibinfo {author} {\bibfnamefont {A.}~\bibnamefont {Hartstein}},\ }\href
  {\doibase 10.1103/PhysRevB.9.3424} {\bibfield  {journal} {\bibinfo  {journal}
  {Phys. Rev. B}\ }\textbf {\bibinfo {volume} {9}},\ \bibinfo {pages} {3424}
  (\bibinfo {year} {1974})}\BibitemShut {NoStop}%
\bibitem [{\citenamefont {Hutchings}\ and\ \citenamefont
  {Van~Stryland}(1992)}]{hutchings1992nondegenerate}%
  \BibitemOpen
  \bibfield  {author} {\bibinfo {author} {\bibfnamefont {D.}~\bibnamefont
  {Hutchings}}\ and\ \bibinfo {author} {\bibfnamefont {E.~W.}\ \bibnamefont
  {Van~Stryland}},\ }\href@noop {} {\bibfield  {journal} {\bibinfo  {journal}
  {JOSA B}\ }\textbf {\bibinfo {volume} {9}},\ \bibinfo {pages} {2065}
  (\bibinfo {year} {1992})}\BibitemShut {NoStop}%
\bibitem [{\citenamefont {Hayat}\ \emph {et~al.}(2008)\citenamefont {Hayat},
  \citenamefont {Ginzburg},\ and\ \citenamefont
  {Orenstein}}]{hayat2008observation}%
  \BibitemOpen
  \bibfield  {author} {\bibinfo {author} {\bibfnamefont {A.}~\bibnamefont
  {Hayat}}, \bibinfo {author} {\bibfnamefont {P.}~\bibnamefont {Ginzburg}}, \
  and\ \bibinfo {author} {\bibfnamefont {M.}~\bibnamefont {Orenstein}},\
  }\href@noop {} {\bibfield  {journal} {\bibinfo  {journal} {Nature photonics}\
  }\textbf {\bibinfo {volume} {2}},\ \bibinfo {pages} {238} (\bibinfo {year}
  {2008})}\BibitemShut {NoStop}%
\bibitem [{\citenamefont {Pines}\ \emph {et~al.}(1978)\citenamefont {Pines},
  \citenamefont {Vega},\ and\ \citenamefont {Mehring}}]{pines-prb-78}%
  \BibitemOpen
  \bibfield  {author} {\bibinfo {author} {\bibfnamefont {A.}~\bibnamefont
  {Pines}}, \bibinfo {author} {\bibfnamefont {S.}~\bibnamefont {Vega}}, \ and\
  \bibinfo {author} {\bibfnamefont {M.}~\bibnamefont {Mehring}},\ }\href
  {\doibase 10.1103/PhysRevB.18.112} {\bibfield  {journal} {\bibinfo  {journal}
  {Phys. Rev. B}\ }\textbf {\bibinfo {volume} {18}},\ \bibinfo {pages} {112}
  (\bibinfo {year} {1978})}\BibitemShut {NoStop}%
\bibitem [{\citenamefont {Carter}\ \emph {et~al.}(2015)\citenamefont {Carter},
  \citenamefont {Soykal}, \citenamefont {Dev}, \citenamefont {Economou},\ and\
  \citenamefont {Glaser}}]{carter-prb-15}%
  \BibitemOpen
  \bibfield  {author} {\bibinfo {author} {\bibfnamefont {S.~G.}\ \bibnamefont
  {Carter}}, \bibinfo {author} {\bibfnamefont {O.~O.}\ \bibnamefont {Soykal}},
  \bibinfo {author} {\bibfnamefont {P.}~\bibnamefont {Dev}}, \bibinfo {author}
  {\bibfnamefont {S.~E.}\ \bibnamefont {Economou}}, \ and\ \bibinfo {author}
  {\bibfnamefont {E.~R.}\ \bibnamefont {Glaser}},\ }\href {\doibase
  10.1103/PhysRevB.92.161202} {\bibfield  {journal} {\bibinfo  {journal} {Phys.
  Rev. B}\ }\textbf {\bibinfo {volume} {92}},\ \bibinfo {pages} {161202(R)}
  (\bibinfo {year} {2015})}\BibitemShut {NoStop}%
\bibitem [{\citenamefont {Simin}\ \emph {et~al.}(2016)\citenamefont {Simin},
  \citenamefont {Soltamov}, \citenamefont {Poshakinskiy}, \citenamefont
  {Anisimov}, \citenamefont {Babunts}, \citenamefont {Tolmachev}, \citenamefont
  {Mokhov}, \citenamefont {Trupke}, \citenamefont {Tarasenko}, \citenamefont
  {Sperlich}, \citenamefont {Baranov}, \citenamefont {Dyakonov},\ and\
  \citenamefont {Astakhov}}]{simin-prx-16}%
  \BibitemOpen
  \bibfield  {author} {\bibinfo {author} {\bibfnamefont {D.}~\bibnamefont
  {Simin}}, \bibinfo {author} {\bibfnamefont {V.~A.}\ \bibnamefont {Soltamov}},
  \bibinfo {author} {\bibfnamefont {A.~V.}\ \bibnamefont {Poshakinskiy}},
  \bibinfo {author} {\bibfnamefont {A.~N.}\ \bibnamefont {Anisimov}}, \bibinfo
  {author} {\bibfnamefont {R.~A.}\ \bibnamefont {Babunts}}, \bibinfo {author}
  {\bibfnamefont {D.~O.}\ \bibnamefont {Tolmachev}}, \bibinfo {author}
  {\bibfnamefont {E.~N.}\ \bibnamefont {Mokhov}}, \bibinfo {author}
  {\bibfnamefont {M.}~\bibnamefont {Trupke}}, \bibinfo {author} {\bibfnamefont
  {S.~A.}\ \bibnamefont {Tarasenko}}, \bibinfo {author} {\bibfnamefont
  {A.}~\bibnamefont {Sperlich}}, \bibinfo {author} {\bibfnamefont {P.~G.}\
  \bibnamefont {Baranov}}, \bibinfo {author} {\bibfnamefont {V.}~\bibnamefont
  {Dyakonov}}, \ and\ \bibinfo {author} {\bibfnamefont {G.~V.}\ \bibnamefont
  {Astakhov}},\ }\href {\doibase 10.1103/PhysRevX.6.031014} {\bibfield
  {journal} {\bibinfo  {journal} {Phys. Rev. X}\ }\textbf {\bibinfo {volume}
  {6}},\ \bibinfo {pages} {031014} (\bibinfo {year} {2016})}\BibitemShut
  {NoStop}%
\bibitem [{\citenamefont {Soltamov}\ \emph {et~al.}(2012)\citenamefont
  {Soltamov}, \citenamefont {Soltamova}, \citenamefont {Baranov},\ and\
  \citenamefont {Proskuryakov}}]{soltamov-prl-12}%
  \BibitemOpen
  \bibfield  {author} {\bibinfo {author} {\bibfnamefont {V.~A.}\ \bibnamefont
  {Soltamov}}, \bibinfo {author} {\bibfnamefont {A.~A.}\ \bibnamefont
  {Soltamova}}, \bibinfo {author} {\bibfnamefont {P.~G.}\ \bibnamefont
  {Baranov}}, \ and\ \bibinfo {author} {\bibfnamefont {I.~I.}\ \bibnamefont
  {Proskuryakov}},\ }\href {\doibase 10.1103/PhysRevLett.108.226402} {\bibfield
   {journal} {\bibinfo  {journal} {Phys. Rev. Lett.}\ }\textbf {\bibinfo
  {volume} {108}},\ \bibinfo {pages} {226402} (\bibinfo {year}
  {2012})}\BibitemShut {NoStop}%
\bibitem [{\citenamefont {Carbonera}(2009)}]{carbonera2009optically}%
  \BibitemOpen
  \bibfield  {author} {\bibinfo {author} {\bibfnamefont {D.}~\bibnamefont
  {Carbonera}},\ }\href@noop {} {\bibfield  {journal} {\bibinfo  {journal}
  {Photosynthesis research}\ }\textbf {\bibinfo {volume} {102}},\ \bibinfo
  {pages} {403} (\bibinfo {year} {2009})}\BibitemShut {NoStop}%
\bibitem [{\citenamefont {Suter}(2020)}]{mr-1-115-2020}%
  \BibitemOpen
  \bibfield  {author} {\bibinfo {author} {\bibfnamefont {D.}~\bibnamefont
  {Suter}},\ }\href {\doibase 10.5194/mr-1-115-2020} {\bibfield  {journal}
  {\bibinfo  {journal} {Magnetic Resonance}\ }\textbf {\bibinfo {volume} {1}},\
  \bibinfo {pages} {115} (\bibinfo {year} {2020})}\BibitemShut {NoStop}%
\bibitem [{\citenamefont {Suter}\ and\ \citenamefont
  {Jelezko}(2017)}]{SUTER201750}%
  \BibitemOpen
  \bibfield  {author} {\bibinfo {author} {\bibfnamefont {D.}~\bibnamefont
  {Suter}}\ and\ \bibinfo {author} {\bibfnamefont {F.}~\bibnamefont
  {Jelezko}},\ }\href {\doibase https://doi.org/10.1016/j.pnmrs.2016.12.001}
  {\bibfield  {journal} {\bibinfo  {journal} {Progress in Nuclear Magnetic
  Resonance Spectroscopy}\ }\textbf {\bibinfo {volume} {98-99}},\ \bibinfo
  {pages} {50} (\bibinfo {year} {2017})}\BibitemShut {NoStop}%
\bibitem [{\citenamefont {Bathen}\ and\ \citenamefont {Vines}(2021)}]{bathen}%
  \BibitemOpen
  \bibfield  {author} {\bibinfo {author} {\bibfnamefont {M.~E.}\ \bibnamefont
  {Bathen}}\ and\ \bibinfo {author} {\bibfnamefont {L.}~\bibnamefont {Vines}},\
  }\href {\doibase https://doi.org/10.1002/qute.202100003} {\bibfield
  {journal} {\bibinfo  {journal} {Advanced Quantum Technologies}\ }\textbf
  {\bibinfo {volume} {4}},\ \bibinfo {pages} {2100003} (\bibinfo {year}
  {2021})}\BibitemShut {NoStop}%
\bibitem [{\citenamefont {Abragam}(1961)}]{abragam-book}%
  \BibitemOpen
  \bibfield  {author} {\bibinfo {author} {\bibfnamefont {A.}~\bibnamefont
  {Abragam}},\ }\href@noop {} {\emph {\bibinfo {title} {The principles of
  nuclear magnetism}}}\ (\bibinfo  {publisher} {Oxford University Press},\
  \bibinfo {address} {UK},\ \bibinfo {year} {1961})\BibitemShut {NoStop}%
\bibitem [{\citenamefont {Nagy}\ \emph {et~al.}(2019)\citenamefont {Nagy},
  \citenamefont {Niethammer}, \citenamefont {Widmann}, \citenamefont {Chen},
  \citenamefont {Udvarhelyi}, \citenamefont {Bonato}, \citenamefont {Hassan},
  \citenamefont {Karhu}, \citenamefont {Ivanov}, \citenamefont {Son} \emph
  {et~al.}}]{nagy2019high}%
  \BibitemOpen
  \bibfield  {author} {\bibinfo {author} {\bibfnamefont {R.}~\bibnamefont
  {Nagy}}, \bibinfo {author} {\bibfnamefont {M.}~\bibnamefont {Niethammer}},
  \bibinfo {author} {\bibfnamefont {M.}~\bibnamefont {Widmann}}, \bibinfo
  {author} {\bibfnamefont {Y.-C.}\ \bibnamefont {Chen}}, \bibinfo {author}
  {\bibfnamefont {P.}~\bibnamefont {Udvarhelyi}}, \bibinfo {author}
  {\bibfnamefont {C.}~\bibnamefont {Bonato}}, \bibinfo {author} {\bibfnamefont
  {J.~U.}\ \bibnamefont {Hassan}}, \bibinfo {author} {\bibfnamefont
  {R.}~\bibnamefont {Karhu}}, \bibinfo {author} {\bibfnamefont {I.~G.}\
  \bibnamefont {Ivanov}}, \bibinfo {author} {\bibfnamefont {N.~T.}\
  \bibnamefont {Son}},  \emph {et~al.},\ }\href@noop {} {\bibfield  {journal}
  {\bibinfo  {journal} {Nature communications}\ }\textbf {\bibinfo {volume}
  {10}},\ \bibinfo {pages} {1954} (\bibinfo {year} {2019})}\BibitemShut
  {NoStop}%
\bibitem [{\citenamefont {Sosnovsky}\ and\ \citenamefont
  {Ivanov}(2021)}]{Sosnovsky-prb-21}%
  \BibitemOpen
  \bibfield  {author} {\bibinfo {author} {\bibfnamefont {D.~V.}\ \bibnamefont
  {Sosnovsky}}\ and\ \bibinfo {author} {\bibfnamefont {K.~L.}\ \bibnamefont
  {Ivanov}},\ }\href {\doibase 10.1103/PhysRevB.103.014403} {\bibfield
  {journal} {\bibinfo  {journal} {Phys. Rev. B}\ }\textbf {\bibinfo {volume}
  {103}},\ \bibinfo {pages} {014403} (\bibinfo {year} {2021})}\BibitemShut
  {NoStop}%
\end{thebibliography}
\end{document}